\documentclass[letterpaper,titlepage,11pt]{article}

\pdfoutput=1

\usepackage{amsmath,amssymb,amsthm,mathrsfs,bbm}
\usepackage{latexsym,amscd,amsbsy,amsfonts,dsfont}
\usepackage{graphicx}
\usepackage[
      colorlinks=true,
      linktocpage,
      linkcolor=blue,
      urlcolor=blue,
      filecolor=black,
      citecolor=red,
      pdfstartview=FitV,
      pdftitle={},
        pdfauthor={Tomas Andrade, Roberto Emparan, Aron Jansen, David Licht, Raimon Luna, Ryotaku Suzuki},
        pdfsubject={},
        pdfkeywords={},
        pdfpagemode=None,
        bookmarksopen=true
      ]{hyperref}\usepackage{caption}

\usepackage{bm}
\usepackage{latexsym}
\usepackage{amssymb}
\usepackage{amsmath}
\usepackage{subfigure}

\usepackage[utf8]{inputenc}

\newcommand{\beq}{\begin{equation}}
\newcommand{\eeq}{\end{equation}}
\newcommand{\beqa}{\begin{eqnarray}}
\newcommand{\eeqa}{\end{eqnarray}}
\newcommand{\bea}{\begin{eqnarray}}
\newcommand{\eea}{\end{eqnarray}}

\newcommand{\nn}{\nonumber}

\newcommand{\ie}{{i.e.,\ }}
\newcommand{\eg}{{e.g.,\ }}

\newcommand{\lp}{\left(}
\newcommand{\rp}{\right)}

\newcommand{\mc}[1]{\mathcal{#1}}
\newcommand{\ord}[1]{{\mathcal O}\lp #1\rp}

\newcommand{\cR}{\mathcal{R}}

\newcommand*{\vcenteredhbox}[1]{\begin{tabular}{@{}c@{}}#1\end{tabular}}

\def\clock{{\count0=\time
           \divide\count0 60
           \ifnum\count0<10 0\fi\the\count0
           \multiply\count0 -60 \advance\count0 \time
           :\ifnum\count0<10 0\fi \the\count0
         }}
\newcommand{\timestamp}{{\small\vbox{\hbox{\tt\jobname.tex}
\hbox{\the\day/\the\month/\the\year, \clock}}}}

\newcommand{\fq}{\mathfrak{q}}

\newcommand{\fr}[1]{\frac{1}{#1}}

\newcommand{\nonum}{\nonumber\\ }

\newcommand{\cout}[1]{}

\numberwithin{equation}{section}

\begin{document}

\begin{titlepage}
\leftline{}
\vskip 1.5cm
\centerline{\LARGE \bf Entropy production and entropic attractors}
\bigskip
\centerline{\LARGE \bf in black hole fusion and fission} 

\vskip 1.2cm
\centerline{\bf Tom{\'a}s Andrade$^{a}$, Roberto Emparan$^{a,b}$, Aron Jansen$^{a}$, 
}
\smallskip
\centerline{\bf 
David Licht$^{a}$, Raimon Luna$^{a}$ and Ryotaku Suzuki$^{a,c}$}

\vskip 0.5cm
\centerline{\sl $^{a}$Departament de F{\'\i}sica Qu\`antica i Astrof\'{\i}sica, Institut de
Ci\`encies del Cosmos,}
\centerline{\sl  Universitat de
Barcelona, Mart\'{\i} i Franqu\`es 1, E-08028 Barcelona, Spain}
\smallskip
\centerline{\sl $^{b}$Instituci\'o Catalana de Recerca i Estudis
Avan\c cats (ICREA)}
\centerline{\sl Passeig Llu\'{\i}s Companys 23, E-08010 Barcelona, Spain}
\smallskip
\centerline{\sl $^{c}$Department of Physics, Osaka City University,}
\centerline{\sl Sugimoto 3-3-138, Osaka 558-8585, Japan}
\smallskip
\vskip 0.5cm
\centerline{\small\tt tandrade@icc.ub.edu,\, emparan@ub.edu, a.p.jansen@icc.ub.edu, } 
\smallskip
\centerline{\small\tt david.licht@icc.ub.edu,\, raimonluna@icc.ub.edu,\, s.ryotaku@icc.ub.edu}

\vskip 1.cm
\centerline{\bf Abstract} \vskip 0.2cm \noindent

\noindent We study how black hole entropy is generated and the role it plays in several highly dynamical processes: the decay of unstable black strings and ultraspinning black holes; the fusion of two rotating black holes; and the subsequent fission of the merged system into two black holes that fly apart (which can occur in dimension $D\geq 6$, with a mild violation of cosmic censorship). Our approach uses the effective theory of black holes at $D\to\infty$, but we expect our main conclusions to hold at finite $D$. Black hole fusion is highly irreversible, while fission, which follows the pattern of the decay of black strings, generates comparatively less entropy. In $2\to 1\to 2$ black hole collisions an intermediate, quasi-thermalized state forms that then fissions. This intermediate state erases much of the memory of the initial states and acts as an attractor funneling the evolution of the collision towards a small subset of outgoing parameters, which is narrower the closer the total angular momentum is to the critical value for fission. Entropy maximization provides a very good guide for predicting the final outgoing states.
Along our study, we  clarify how entropy production and irreversibility appear in the large $D$ effective theory.  We also extend the study of the stability of new black hole phases (black bars and dumbbells). Finally, we discuss entropy production through charge diffusion in collisions of charged black holes.
\end{titlepage}
\pagestyle{empty}
\small

\addtocontents{toc}{\protect\setcounter{tocdepth}{2}}

\tableofcontents
\normalsize
\newpage
\pagestyle{plain}
\setcounter{page}{1}

\section{Introduction and Summary}

The area theorem, or second law of black holes, has pervasive implications in all of black hole physics. It puts absolute bounds on gravitational wave emission in collisions (Hawking's original motivation in \cite{Hawking:1971tu}) and limits other classical black hole evolutions, but also, through the identification of horizon area as entropy \cite{Bekenstein:1974ax}, it gives an entry into quantum gravity, holography, and applications of the latter to strongly coupled systems.

Investigating the growth of black hole entropy should throw interesting light into complex dynamical black hole processes. How does the second law constrain the possible final states? Are there phenomena where it can provide more than bounds on allowed outcomes, for instance, indicating their likelihood, according to how much entropy they generate? Since the area of the event horizon can be computed outside stationary equilibrium, one may even study the mechanisms that drive its growth at different stages.

Unfortunately, computing this entropy during a highly dynamical process, such as a black hole merger, is in general very complicated and requires sophisticated numerical calculations. In this article we resort to an approach that simplifies enormously the task: the effective theory of black holes in the limit of a large number of dimensions, $D\to\infty$ \cite{Emparan:2013moa,Emparan:2020vyfinr}, developed in \cite{Emparan:2015hwa,Bhattacharyya:2015dva,Emparan:2015gva,Bhattacharyya:2015fdk}. We use the equations of \cite{Emparan:2015gva,Emparan:2016sjk}  for the study of asymptotically flat black holes, their stability, and collisions between them \cite{Andrade:2018nsz,Andrade:2018yqu,Andrade:2019edf}.

We examine in detail the production of entropy---its total increase, but also its generation localized in time and in space on the horizon---in processes where an unstable black hole (a black string \cite{Gregory:1993vy,Lehner:2010pn} or an ultraspinning black hole \cite{Myers:1986un,Emparan:2003sy}) decays and fissions, and in collisions where two black holes fuse into a single horizon.  If the total angular momentum in the collision is not too large, the fusion ends on a stable rotating black hole. However, as shown in \cite{Andrade:2018yqu,Andrade:2019edf}, when $D$ is large and if the total angular momentum is also large enough, the merger does not end in the fusion, but proceeds to fission: the intermediate merged horizon is unstable, pinches at a neck---in a mild violation of cosmic censorship \cite{Andrade:2019edf,Emparan:2020vyf}---and arguably breaks up into two (or possibly more) black holes that then fly apart. 
The phenomenon, up until the formation of the singular pinch, has been verified to occur in $D=6,7$ through numerical analysis in \cite{Andrade:2020}. The importance of the intermediate phase in the evolution of the system was noticed in the earlier studies \cite{Andrade:2018yqu,Andrade:2019edf}, but here we will go significantly further in revealing how it controls the outcome.



Before we present our main results, we shall discuss general issues related to area growth, its identification with entropy, and its computation in the large $D$ effective theory.

\paragraph*{Black hole entropy and its growth.} 

In General Relativity the horizon area increases through two effects: the addition of new generators to the horizon (at caustics or crossover points, on spacelike crease sets), and the expansion of pencils of existing generators.  In this article, the methods and approximations that we employ allow to study the latter, \ie how the area expands smoothly. The effective theory of large $D$ black brane dynamics provides explicit entropy production formulas for viscous dissipation on the horizon \cite{Emparan:2016sjk,Herzog:2016hob,Bhattacharyya:2016nhn,Dandekar:2017aiv}, which we apply to the evolution of unstable asymptotically flat black holes, and to the fusion and fission in black hole collisions. 
The addition of generators through caustics is actually suppressed when $D$ is large\footnote{This can be seen, for instance, in extreme-mass-ratio mergers with the methods of \cite{Emparan:2016ylg}.}. We expect that, at finite $D$, the entropy growth through this addition in the merger of two black holes is important only during the first instants of the fusion, and much less so during the relaxation.   In fission, the break-up of the horizon across a naked but mild singularity involves a region of very small horizon area\footnote{A Planckian area, which vanishes in the classical limit.} and therefore loses only a few generators in any dimension, and even fewer as $D$ grows large. This process is controlled by quantum gravity, but we have argued elsewhere that the effects upon the classical evolution should be negligibly small \cite{Emparan:2020vyf}.

The notions of entropy and entropy current that we use are associated to the area of the event horizon, and in this respect they are closely related to the ones in the Fluid/Gravity correspondence in \cite{Bhattacharyya:2008jc,Bhattacharyya:2008xc}. Indeed, we expect that the discussion in that context carries over to the large $D$ formulation: it is possible that other currents with non-negative divergence can be constructed. A related concern is whether one should identify the entropy with the area of the apparent horizon, as there are general arguments in favor of this \cite{Engelhardt:2017aux}, and it has been possible to identify corresponding currents in Fluid/Gravity \cite{Booth:2011qy}. While it would be interesting to further investigate this in the large $D$ effective theory, we will not be concerned with it here. The divergence of the entropy current that we use is as expected for a physical fluid (from viscous dissipation of shear and expansion of the fluid), so at the very least our results will not be unreasonable. Moreover, we expect that the growth properties of different entropy notions will be very similar. Within the large $D$ effective theory, the system evolves smoothly and continuously and so we expect the entropy to do that too. With these methods one does not capture the less smooth features (\eg caustics) of the event horizon in the first stages of the merger, and large discontinuous jumps in the area of the apparent horizon are not expected; these should be suppressed when $D\gg 1$. So, despite the ambiguities in the definition of out-of-equilibrium entropy, we expect that our conclusions remain qualitatively valid for other viable notions of it. 

The other main aspect of finite $D$ physics that is not captured by our methods is the production of gravitational waves, which implies that in our calculations the total energy and angular momentum of the black holes are conserved, making it easier for us to characterize the evolutions. Again, we expect that this radiation is stronger during the initial instants when the black holes first come together. Radiation effects should quickly become less relevant as the number of dimensions grows \cite{Cardoso:2002pa,Emparan:2013moa,Bhattacharyya:2016nhn,Andrade:2019edf}.

The upshot is that we expect that the patterns of entropy production that we find are broadly applicable in $D\geq 6$, and possibly even qualitatively valid for fusion in $D=4$. We will return to this last point near the end.

\paragraph*{Main conclusions.}

All the collisions we study are symmetric, \ie between black holes of equal mass and spin.\footnote{In $2\to 1\to 2$ collisions, the two outgoing black holes will both have the same mass and spin, but the initial and final spins will in general be different. Mass conservation to leading order in $1/D$ implies that the final masses are the same as the initial ones.} This is mostly for simplicity; our methods allow to collide black holes with generic parameters.

\begin{figure}[t]
\centering
 \vcenteredhbox{\includegraphics[width=0.55\linewidth]{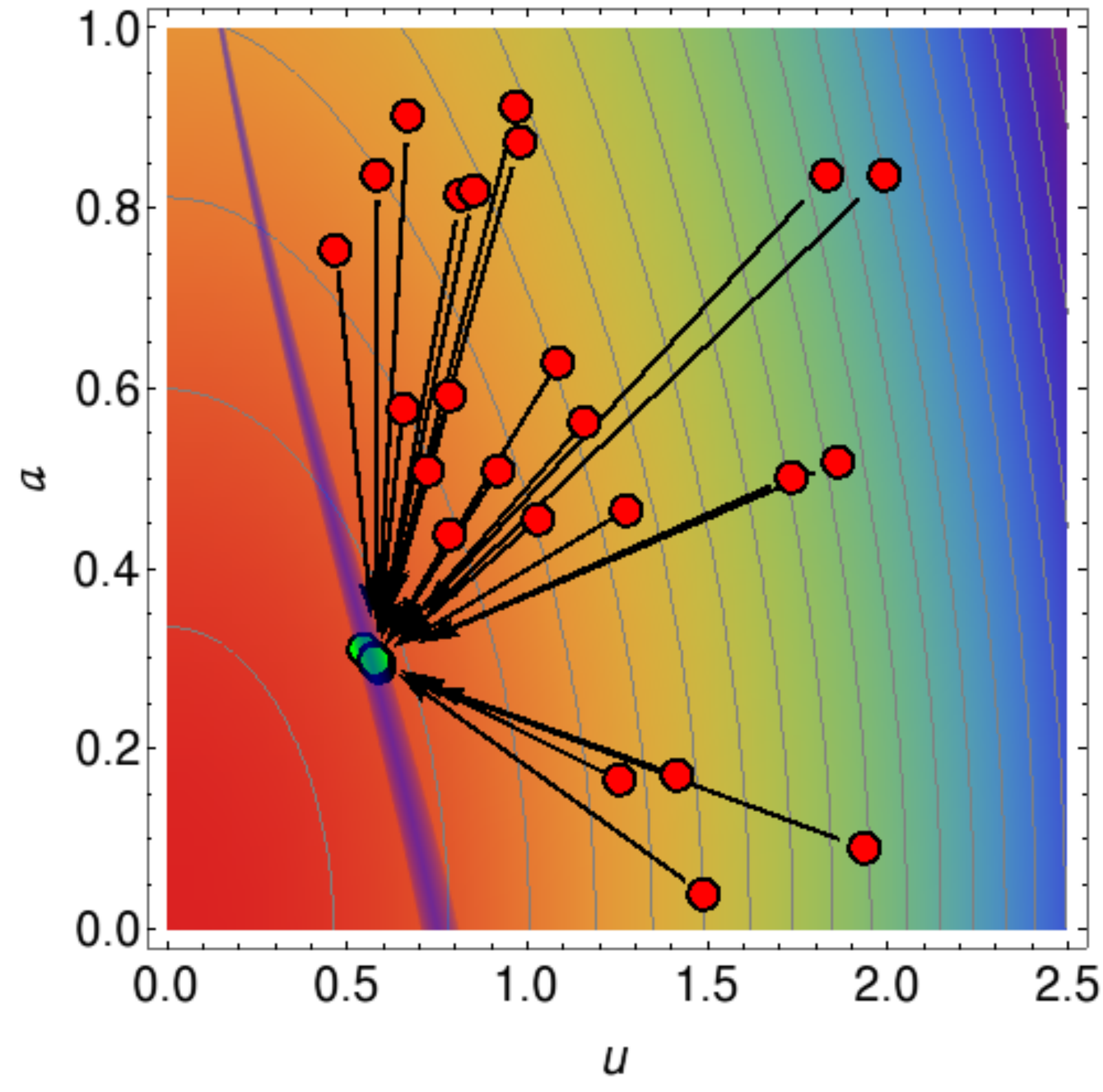}}
\hspace{1em}
 \vcenteredhbox{\includegraphics[width=0.4\linewidth]{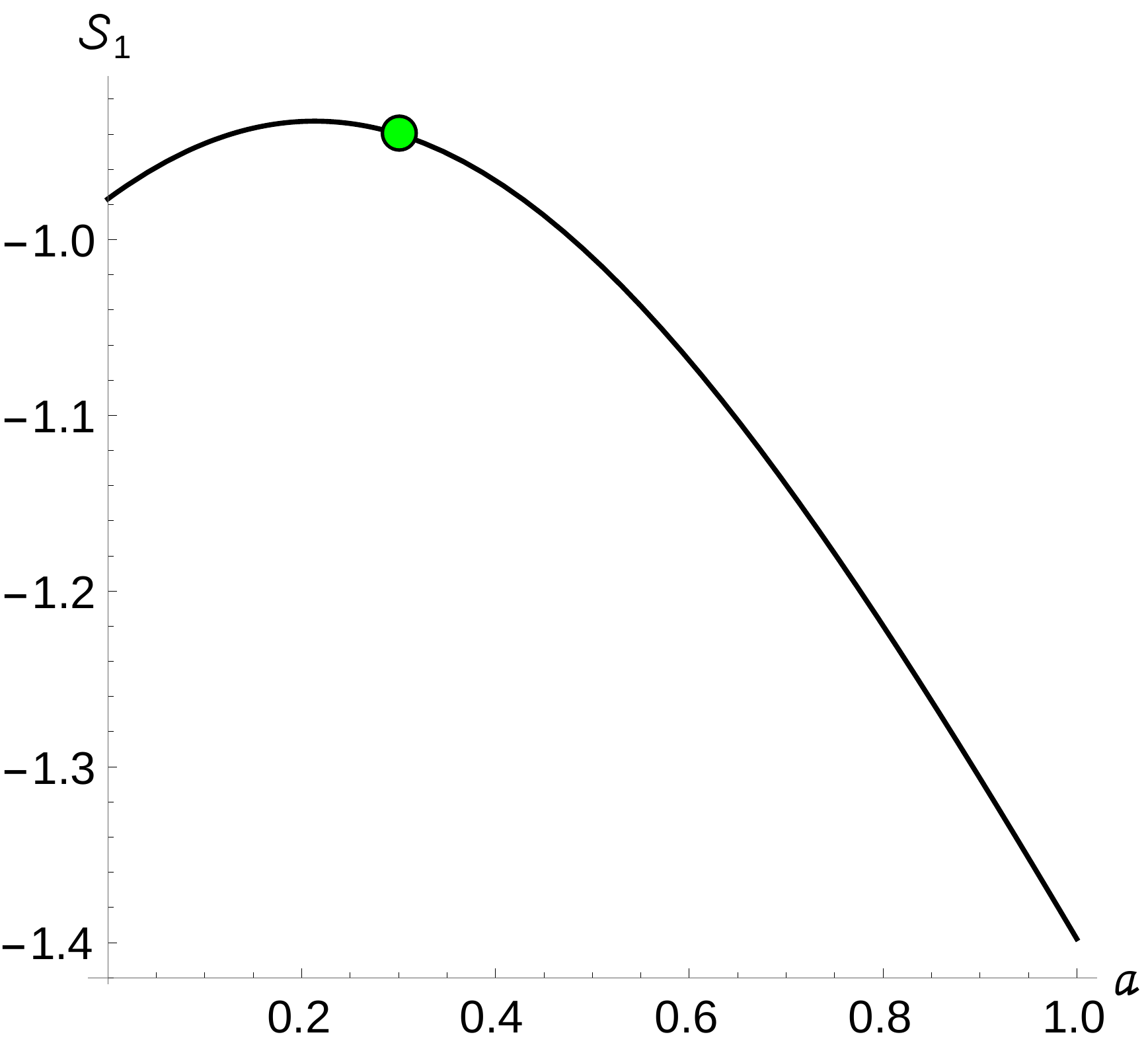}}
\caption{\small Entropic attractor in black hole collisions $2\to 1\to 2$. \emph{Left}: red dots represent initial states, connected by an arrow line to the corresponding  final states (green) after dynamical evolution. The initial and final pairs of black holes are characterized by their rotation (spin) parameter $a$ and their linear velocity $u$ (ingoing or outgoing).  The (conserved) total angular momentum per unit mass in these collisions is fixed to $J/M=2.8$. We see that, independently of the initial states, the final states cluster on $(a,u)\approx (0.3,0.6)$. This is due to the formation of an intermediate, long-lived, quasi-thermalized phase. The contour colors correspond to the (NLO) entropy $\mc{S}_1$ of the configuration. Entropy would be maximized at the lower-left corner, but this would correspond to infinite impact parameter $b$, which is unphysical since $b$ is constrained by the geometric size of the collision. 
The thin purple strip is the region where $b$ takes on geometrically-allowed values for final states. \emph{Right}: entropy along the central value of the purple strip. The attractor is close to the maximum possible final entropy; larger values of $a$ (smaller values of $u$) would be entropically disfavoured, while in the opposite direction the entropy gain would be very small.
\label{fig:attract}}
\end{figure}

Our analysis shows that:
\begin{itemize}

\item Black hole fusion generates comparatively much more entropy, and at faster rates, than black hole fission.

\item Unstable black strings decay with a simple pattern of entropy production which is reproduced in other fission processes.


\item Merger collisions of two black holes have a critical value of the total angular momentum per unit total mass\footnote{These $J$ and $M$ are defined in the effective theory; the corresponding physical quantities are given in appendix~\ref{app:phys}.} 
\beq\label{JMc}
\lp\frac{J}{M}\rp_c\approx 2.66
\eeq
that divides  low $J/M$ collisions $2\to 1$ that end in fusion, from higher $J/M$ collisions $2\to 1\to N\geq 2$ that evolve to fission. The bound holds except for `grazing' mergers with large initial impact parameters.


\item $2\to 1\to 2$ collisions are dominated by the formation of intermediate, long-lived, quasi-stationary, bar-like entropic attractors (fig.~\ref{fig:attract}):

{\renewcommand{\labelenumi}{(\alph{enumi})}
\begin{itemize}

\item The intermediate quasi-thermalization largely erases the memory of the initial state, so the final outgoing states are almost independent of the initial parameters, other than the total conserved $J/M$.

\item This attractor effect is stronger the closer $J/M$ is to the critical value \eqref{JMc}.

\item The  attractor can be approximately (but not exactly) predicted by maximizing the entropy generation among possible outgoing black holes.

\end{itemize}
}

\item Entropy is produced through viscous dissipation of shear and expansion of the effective velocity field. In fusion generically, and in fission always, both enter in almost equal proportion. The formation of the intermediate bar phase can be dominated by shearing depending on initial conditions.

\end{itemize}

The reason the attractor effect is stronger when $J/M$ is near criticality is that the intermediate state is closer to a marginally stable solution. When $J/M$ is higher, the intermediate black hole is shorter-lived and its features are less precisely defined, so its decay outcomes show larger spread. We emphasize that the attractor is a feature of $2\to 1\to 2$ collisions with intermediate fusion; this requires that the initial impact parameter and initial velocities are not too large, otherwise the two black holes fly by each other. Fusionless $2\to 2$ collisions are not included in fig.~\ref{fig:attract}, and are little studied in this article.

Let us elaborate on the near-maximization of entropy in collisions $2\to 1\to 2$. After fixing an overall scale by setting the total mass to one, the outgoing black holes are characterized by their spin parameter $a$, outgoing velocity $u$, and outgoing impact parameter $b$.\footnote{There is one more outgoing parameter: the scattering angle. However, this is not affected by conservation laws nor by entropic considerations, and we shall have little to say about it.} We argue that these can be well predicted by considering three different constraints:
{\renewcommand{\labelenumi}{(\roman{enumi})}
\begin{enumerate}
\item Kinematic: the total final angular momentum must be the same as the initial one, which imposes a relation between $a$, $b$ and $u$.
\item Geometric: the outgoing impact parameter is limited by the geometrical size of the collision. This yields a constraint between $b$ and $a$ (appendix~\ref{app:bout}).
\item Entropic: after imposing (i) and (ii), near-maximization of the final entropy gives a good approximation to the final values of $a$, $b$, and $u$.
\end{enumerate}
}
In fig.~\ref{fig:attract}, (i) is included by considering states with given total $J/M=2.8$, while (ii) restricts final states to the lie along the purple band; the graph on the right then makes (iii) apparent.

The most remarkable of these constraints is entropy maximization: it provides a simple proxy for the complex dynamics that drives the system to its final state. We do not have an answer to why it is not a perfect predictor---other than there is no reason that it should be---but given our results it is natural to wonder how accurate it becomes as $J/M$ approaches from above the critical value \eqref{JMc}. 

The principle of entropy maximization actually holds quite well in four-dimensional black hole collisions: in the merger, the final entropy would be maximal (consistent with conservation of energy and angular momentum) if no gravitational waves were emitted. The fact that the radiated energy is typically only a few-percent fraction of the total energy means that the final entropy is only a few percent off the maximum. In the limit $D\to\infty$, the $2\to 1$ fusion trivially maximizes the entropy since radiation is absent. The fission of an unstable black object, instead, has a range of possible outcomes. For instance, black strings can split up into several blobs, and the final entropy is larger for fewer blobs. The decay of ultraspinning black holes (MP, bars, and dumbbells) is more similar to the fission stage in $2\to 1\to 2$ collisions, but the evolution of the instability is sensitive to the specific perturbation that triggers it. The process starts with an unstable system and looks more contrived and less natural than a collision, so in our study we have focused mostly on the latter. Note, however, that the decay of the critical, marginally stable solution at \eqref{JMc} may illuminate the question of how closely can entropy be maximized. This deserves closer examination.

\paragraph*{Entropy generation and irreversibility in the leading order (LO) large $D$ effective theory.}

Readers familiar with the large $D$ effective theory of black holes and branes may be surprised that the entropy growth can be computed with its equations to leading order in the $1/D$ expansion. This theory is known to exactly conserve the entropy of the system: the LO entropy current is divergence-free, and entropy generation is suppressed by a factor $1/D$ \cite{Emparan:2015gva,Bhattacharyya:2016nhn}. A simple illustration of this property is the fusion of two equal mass Schwarzschild black holes \cite{Caldarelli:2008mv,Emparan:2013moa}, each with entropy
\beq\label{bhentD}
S(M)\propto M^{\frac{D-2}{D-3}}\,,
\eeq
which merge into a single one, so that (since losses into radiation are suppressed non-perturbatively in $1/D$)
\beq\label{fusionent}
\frac{S_\text{final}}{S_\text{initial}}=\frac{S(2M)}{2S(M)}=1+\frac{\ln 2}{D} +\ord{\frac1{D^2}}\,,
\eeq
\ie the entropy increase is $\propto 1/D$. This feature extends to all of the dynamics of black branes at large $D$ described by the LO effective theories of \cite{Emparan:2015gva,Bhattacharyya:2016nhn}.

This seems to make it impossible to see entropy growth unless one employs the next-to-leading order (NLO) theory. It also raises a puzzle: if the LO entropy does not grow, how can we characterize the irreversibility of the evolution in the LO theory?

We will argue that there exists a quantity $S_1(t)$ in the LO theory such that the evolution equations imply $\partial_t S_1(t)\geq 0$, and hence characterizes the irreversibility of this theory. This $S_1$ is actually the NLO ($1/D$ suppressed) entropy density, but we might not (indeed, need not) have known this, since $S_1$ and its variations are all given by LO magnitudes. 

The argument of \eqref{fusionent} is still valid when the black holes rotate, since rotation effects in the entropy are suppresed by $1/D$ \cite{Emparan:2015gva,Andrade:2018nsz}. However, it ceases to apply if the black holes carry charge. Correspondingly, the effective theory of large $D$ charged black branes \cite{Emparan:2016sjk,Andrade:2018rcx} allows entropy production, through charge diffusion (resistive Joule heating), at leading order in $1/D$.\footnote{However, entropy production in the theory of charged membranes in \cite{Bhattacharyya:2015fdk} is zero at LO.} The study of entropy production in charged collisions is much simpler than in the neutral case, since it is largely dictated by conservation laws, but it is still illustrative and confirms the conclusions above. We will discuss it after our analysis of neutral collisions.

\paragraph{Outline.} In section~\ref{sec:efftheory} we introduce the basic elements of the large $D$ effective theory of black branes, specifically how entropy generation can be studied within the context of the LO theory. In sec.~\ref{sec:bhcoll} we discuss localized black hole solutions in this effective theory and how their stability properties influence the outcome of collisions. In sec.~\ref{sec:entpro} we perform numerical simulations of evolutions of instabilities and collisions. We investigate in detail the generation of entropy, in time and in space, and use it to characterize the different stages in the collision. The study in sec.~\ref{sec:bhscat} of the scattering of black holes reveals the role as an attractor of the intermediate state which nearly maximizes entropy production. Sec.~\ref{sec:chadif} describes how entropy is produced through charge diffusion in collisions between charged black holes. We conclude in sec.~\ref{sec:concl}.

\section{Entropy production in the large $D$ effective theory}\label{sec:efftheory}

We begin with a discussion of the large $D$ effective theory of black branes, with a focus on entropy and its generation. As we will review in the next section, this theory can be used to study localized, asymptotically flat black holes (Schwarzschild-Tangherlini, Myers-Perry, and others) at large $D$, and their collisions. The extension of the results in this section to the large $D$ effective theory of AdS black branes is straightforward (appendix~\ref{app:AdS}).

The field variables of the effective theory of large $D$ black branes are the mass density of the brane $m(t,\mathbf{x})$ and the velocity field $v^i(t,\mathbf{x})$, where $\mathbf{x} = (x^i)$, and $i=1,\dots, p$, label directions on the flat worldvolume of the brane. Only these $p$ spatial directions have non-trivial dynamics, while the remaining 
\beq
n=D-p-3
\eeq
dimensions play a passive role, serving to perform the large-$D$ localization of dynamics near the horizon of the black hole.

We refer to \cite{Emparan:2015gva,Emparan:2016sjk} for how $m(t,\mathbf{x})$ and $v^i(t,\mathbf{x})$ determine the geometry of a black brane in a spacetime with a large number of dimensions $D$. 
In order to have a solution of the Einstein equations in the large $D$ limit, they must solve the effective field equations \cite{Emparan:2015gva,Emparan:2016sjk,Dandekar:2016jrp}
\beq\label{dtm}
\partial_t m+\partial_i\lp m v^i\rp=0\,,
\eeq
\beq\label{dtmv}
\partial_t (m v^i) +\partial_j \lp m v^i v^j+\tau^{ij}\rp
=0\,,
\eeq
with
\beq\label{stressneut}
\tau_{ij}= - m\,\delta_{ij}  -2m \partial_{(i}v_{j)}- m\,\partial_j\partial_i \ln m\,.
\eeq
Written in this form, they are the equations of a non-relativistic, compressible fluid with mass-energy density $m$, velocity $v^i$, and stress tensor $\tau_{ij}$. 
The first two terms in the stress tensor \eqref{stressneut} correspond, respectively, to (negative) pressure
\beq
P=- m\,,
\eeq
and to viscous terms, with shear and bulk viscosities
\beq
\eta=m\,,\qquad 
\zeta=\frac{2}{p}\eta\,.
\eeq
The last term in \eqref{stressneut}, in a hydrodynamic interpretation, is a second-order transport term.\footnote{It is more easily understood in the elastic interpretation of the effective equations as coming from the extrinsic curvature of the brane \cite{Emparan:2016sjk}.} In the large $D$ limit it must not be assumed to be small: it enters at the same order in $1/D$ as the other terms in \eqref{stressneut}.

The entropy density and the temperature can be obtained from the area density and surface gravity of the black brane. They are
\beq\label{sTneut}
s=4\pi m\,,\qquad T=\frac1{4\pi}\,,
\eeq
and they satisfy the expected thermodynamic relations for the system,
\beq
m=Ts\,,\qquad dm=Tds\,.
\eeq

\subsection{Irreversibility in the effective theory}

Let us now address in detail an elementary puzzle of this effective theory that we alluded to in the introduction. The equations \eqref{dtm}, \eqref{dtmv} contain viscous dissipation, which is expected to render the evolution irreversible.\footnote{This is also apparent using the variable $p_i=mv_i +\partial_i m$, in which \eqref{dtm} and \eqref{dtmv} take the form of inhomogeneous heat equations \cite{Emparan:2015gva}.} This is further confirmed by the spectrum of linearized perturbations, which has quasinormal frequencies with imaginary parts. After all, these equations describe horizons, which are dissipative systems par excellence, but for the moment let us forget black holes and regard by itself the system that these effective equations describe.

On very general grounds, we expect that dissipation in a thermodynamic system creates entropy, reflecting the irreversibility of the evolution. However, in the theory described by \eqref{dtm} and \eqref{dtmv}  the total entropy
\beq
S(t)=\int d^p x\, s(t,x)
\eeq
remains constant in time, since  \eqref{sTneut} implies that it is exactly proportional to the total mass and this is conserved by  \eqref{dtm}. 
So, if the entropy is not growing, what, then, characterizes the irreversibility?

Remarkably, we can identify a quantity in this theory that is strictly non-decreasing  under time evolution. Define the density\footnote{The factors $4\pi$ that we carry over have their origin in $T$ in \eqref{sTneut}.}
\beq\label{s1}
s_1=4\pi\lp -\frac12 m v_i v^i -\frac1{2m}\partial_i m\,\partial^i m +m \ln m\rp\,.
\eeq
We will justify the choice presently, but for now note that using the field equations \eqref{dtm} and \eqref{dtmv} it follows that
\beq\label{dts1}
\partial_t s_1 +\partial_i {j_1}^i =8\pi m\lp \partial_{(i} v_{j)}\rp\lp\partial^{(i} v^{j)}\rp\,,
\eeq
where
\beq\label{j1}
{j_1}^i =s_1 v^i -4\pi \lp v^j \tau^{ij}+(\partial_j m)(\partial^j v^i)\rp \,.
\eeq
Since the right hand side of \eqref{dts1} is non-negative, we conclude that
\beq\label{S1}
S_1(t)=\int d^p x\, s_1(t,x)
\eeq
is a non-decreasing function in time,
\beq\label{dtS1}
\partial_t S_1(t)\geq 0\,.
\eeq
This characterizes the irreversibility of the evolution in the effective theory.

Observe that the growth rate
\beqa\label{dtSetazeta}
\partial_t S_1(t) &=& 8\pi \int d^p x\, m \lp \partial_{(i} v_{j)}\rp\lp\partial^{(i} v^{j)}\rp\nn\\
&=&\int d^p x\, \lp \frac{2\eta}{T}\sigma_{ij}\sigma^{ij} +\frac{\zeta}{T}\lp\partial_i v^i\rp^2\rp
\eeqa
is that of a hydrodynamic entropy generated by viscous heating, with contributions from dissipation of shear 
\beq
\sigma_{ij}=\partial_{(i}v_{j)} -\frac1{p}\delta_{ij}\, \partial_k v^k
\eeq
and dissipation of expansion $\partial_i v^i$. This is also a feature of the entropy in the Fluid/Gravity correspondence \cite{Bhattacharyya:2008xc,Dandekar:2017aiv} (see \cite{Emparan:2020vyfinr} for further discussion in these contexts).

\subsection{Entropy at next-to-leading order}\label{subsec:NLOent}

The explanation for these properties of $S_1$ is that it is actually the leading $1/D$ contribution to the black brane entropy. Namely, the entropy density obtained from the event horizon area of the black brane is, up to a total divergence (see appendix~\ref{app:nlo})
\beq\label{nloent}
s(t,x)=4\pi\bar{m}(t,x)\lp 1+\frac{c_s}{D}\rp+\frac1{D} s_1(t,x)\,,
\eeq 
where $\bar{m}$ is the energy density including NLO corrections, and $c_s$ is a constant (which we determine in appendix~\ref{app:phys}) that accounts for the fact that, in order to simplify the form of $s_1$, we have subtracted a term $\propto m$ without changing the right-hand side of \eqref{dts1}. The total entropy is
\beq\label{SS1}
S(t)=4\pi M\lp 1+\frac{c_s}{D}\rp +\frac1{D}S_1(t)\,.
\eeq
Eq.~\eqref{dtSetazeta} then gives the production rate of entropy to NLO in the $1/D$ expansion. The point to notice is that, since the LO entropy is proportional to the energy, which is constant to all orders, the time derivative of the entropy at NLO can be computed using only quantities of the LO effective theory. This is what allows us to identify within this theory the quantities $s_1$ and $S_1$ which behave irreversibly.

Observe that we can write \eqref{nloent} as
\beq\label{nloent2}
s(t,x)=4\pi \lp \bar{m}-\frac1{D} \lp \frac12 m v^2 +\frac1{2m}(\partial m)^2-c_s\rp\rp^{1+1/D}\,,
\eeq
which we can understand as follows. The dependence of entropy on mass  \eqref{bhentD} for a Schwarzschild black hole at finite $D$ is $S\propto M^{1+1/(D-3)}$, which is like \eqref{nloent} at large $D$. The term $-\frac12 m v^2$ is a kinetic energy\footnote{Physical velocities in the effective theory are rescaled by a factor $1/\sqrt{D}$ \cite{Emparan:2015gva}, which explains why the term is $1/D$ suppressed.}. It appears here because, out of the total energy of the black hole, only its rest (irreducible) mass contributes to entropy. Observe that this motion could be linear, as in a boosted black hole, or circular, as in a rotating black hole: both reduce the `heat' fraction of the total energy. The last term in \eqref{nloent2} is (likely) a correction from curvature of the horizon due to the difference between the radial position measured by $m$ and the actual area density.  

The entropy density \eqref{s1} simplifies for stationary solutions which rotate rigidly, such that \cite{Emparan:2016sjk}
\beq
\partial_t m +v^i\partial_i m=0\,,\qquad \partial_t v^i=0\,,\qquad \partial_{(i}v_{j)}=0\,.
\eeq
In this case the effective equations reduce to the `soap bubble equation'
\beq\label{soapbubble}
\frac12 m v^i v_i+m\ln m+\partial_i\partial^i m-\frac1{2m}\partial_i m\,\partial^i m=c\, m\,,
\eeq
where $c$ is an integration constant that corresponds to a choice of scale for the total mass. We set it to zero, since in the end we will work with scale-invariant quantities where $c$ would disappear. Using this equation we obtain, after dropping a boundary term from a total derivative,
\beq
S_1(t)=-4\pi\int d^p x\, m v_i v^i\,.
\eeq
For a solution that rotates along independent angles $\phi_a$ with velocities $v^{\phi_a}=\Omega^a$, this gives
\beq
T S_1= -\Omega^a J_a\,,
\eeq
where $T$ is the LO temperature \eqref{sTneut}. It is easy to verify that, when added to the LO entropy, this reproduces the Smarr relation for black holes at NLO in $1/D$.

\subsection{Measuring the entropy}

When we compare the entropy of different solutions---\eg ingoing and outgoing black holes---we will do it between configurations with the same total mass. For this purpose, if $\mathbf{S}$ and $\mathbf{M}$ are the physical entropy and mass of the black hole in $D$ dimensions, one works with a mass-normalized, scale-invariant, dimensionless entropy of the form
\beq\label{mcS}
\mc{S}=C \frac{\mathbf{S}}{\mathbf{M}^{\frac{D-2}{D-3}}}\,.
\eeq
Here 
\beq
C =\lp \frac{(D-2)\Omega_{D-2}}{16\pi G}\rp ^{1/(D-3)}\frac{D-2}{4\pi}
\eeq
is a suitable convention to simplify later expressions; it could be set to one by adequately choosing Newton's constant $G$. 

Similarly, in the effective theory we define a mass-normalized, scale-invariant entropy,
\beq\label{calS1}
\mc{S}_1(t)=\frac{S_1(t)}{4\pi M}-\ln \frac{M}{2\pi e^2}\,.
\eeq
Subtraction of the term $\ln M$ makes this quantity independent of the choice of mass scale, in particular of the value of $c$ in \eqref{soapbubble}. We have also added a constant $\ln(2\pi e^2)$ to simplify later expressions. One can then verify (see appendix~\ref{app:phys}) that the physical mass-normalized entropy \eqref{mcS} is given in terms of the effective theory one \eqref{calS1} by
\beq\label{physSeftS1}
\mc{S}= 1+ \frac1{D}\mc{S}_1(t)+\ord{\frac1{D^2}}\,.
\eeq

\section{General features of black hole collisions}\label{sec:bhcoll}

In the following we restrict to black holes with rotation on a single plane, which will also be the plane on which the black holes move and collide. Then, in the effective theory we study configurations with non-trivial dependence in only $2+1$ dimensions, \ie on a 2-brane.


\subsection{Brane blobology }

The effective equations have stationary solutions that describe localized black holes rotating with angular velocity $\Omega$ \cite{Andrade:2018nsz} such that 
\beq
v_i=\Omega\, \varepsilon_{ij}x^j\,.
\eeq
The Myers-Perry (MP) black holes are given by 
\beq\label{MPblob}
m(x_1,x_2)=m_0\exp \lp-\frac{x_i x^i}{2(1+a^2)}\rp\,,\qquad \Omega =\frac{a}{1+a^2}\,.
\eeq
The mass is
\beq
M=2\pi m_0 (1+a^2)\,.
\eeq
Here $m_0$ measures the horizon radius at the rotation axis, and with our choice of $c=0$ in \eqref{soapbubble} it is
\beq\label{m0C}
m_0=\exp \lp \frac2{1+a^2}\rp\,.
\eeq
While $M$ depends on the choice of scale of normalization,
more relevant are scale-invariant quantities, namely the spin per unit mass
\beq\label{JMa}
\frac{J}{M}=2a
\eeq
and the NLO mass-normalized entropy
\beq\label{calS1MP}
\mc{S}_1= -\ln (1+a^2)\,.
\eeq
The fact that $\mc{S}_1$ becomes more negative with larger $a$ is the familiar decrease of the black hole entropy as the spin grows, in the large $D$ limit.

Another exact solution describes rotating black bars,
\beq
m= \exp\lp 1-\frac{x_c^2}{4}\lp 1+\sqrt{1-4\Omega^2} \rp-\frac{y_c^2}{4}\lp 1-\sqrt{1-4\Omega^2} \rp \rp\,,
\eeq
where $x_c$, $y_c$ are corotating coordinates given by
\begin{align}
	x_c&=x^{1} \cos \Omega t+x^{2} \sin \Omega t\, , \nn\\
	y_c&=x^{2} \cos \Omega t-x^{2} \sin \Omega t\,,
\end{align}
and our scale normalization is again consistent with $c=0$ in \eqref{soapbubble}. This solution has
\beq
M=\frac{2 \pi} {\Omega}\,,
\eeq
and
\beq
\frac{J}{M}=\Omega^{-1}\,,\qquad
\mc{S}_1=\ln \Omega\,.
\eeq
For $\Omega=1/2$ this branch of solutions joins the MP family.

\begin{figure}[t]
\centering
\includegraphics[width=0.45 \linewidth]{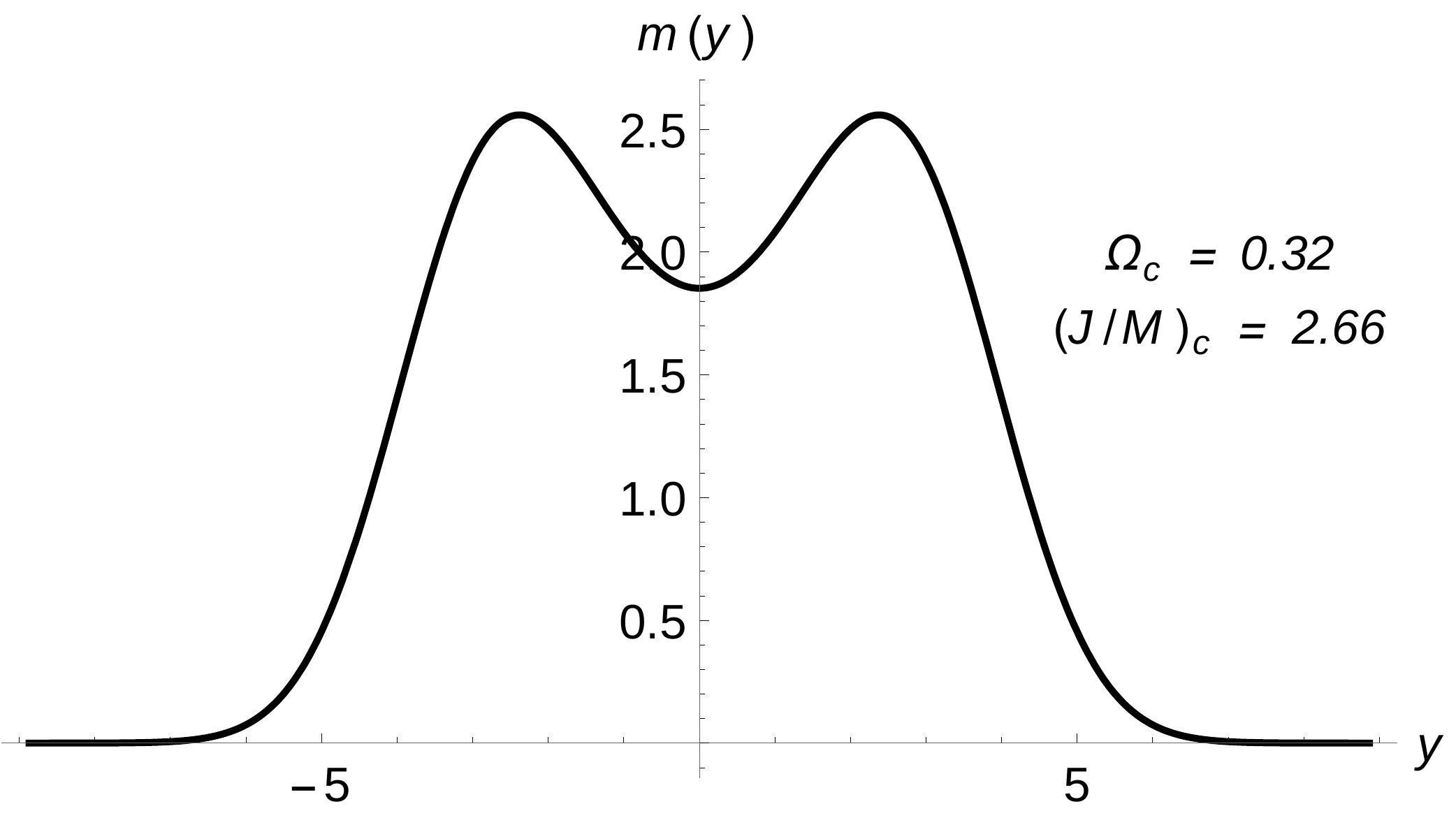}
\caption{\small Profile along the long axis of the `critical' dumbbell solution with $J/M$ equal to \eqref{JMc}. \label{fig:critdumb}}
\end{figure}

Ref.~\cite{Licht:2020odx} constructed numerically large classes of other stationary solutions. The most relevant for us are rotating dumbbells. They can be regarded as black bars with a pinch in their middle, see fig.~\ref{fig:critdumb}.

\subsection{Phase diagrams and the outcomes of collisions}\label{subsec:phases}

\begin{figure}[t]
 \begin{center}
 \includegraphics[width=.7\textwidth]{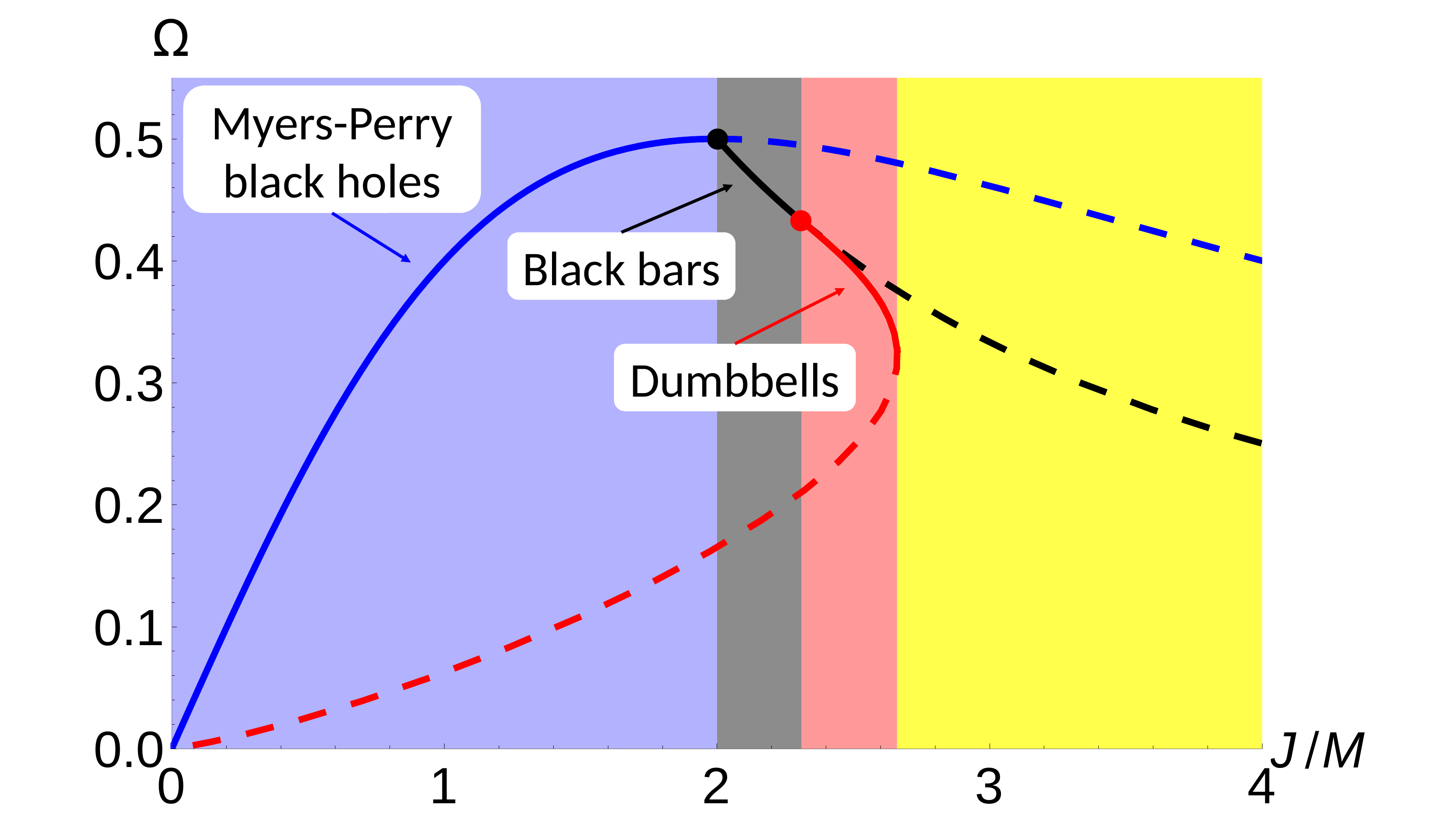}
 \end{center}
 \caption{\small Phases of blobs and their stability as relevant to outcomes of mergers. Solid/dashed lines are stable/unstable stationary blobs. Blue: Myers-Perry black holes, stable up to $J/M=2$. Black: black bars, stable up to $J/M=4/\sqrt{3}\approx 2.31$. Red: black dumbbells, stable up to $J/M=(J/M)_c\approx 2.66$ (dumbbells along the dashed line are unstable binaries of blobs). The background shading indicates the expected outcome of a merger for an initial value of $J/M$. No stable stationary blobs exist for $J/M>(J/M)_c$ (yellow), so, if a merger occurs in this region, it can only evolve to a multi-blob state. The numerical solutions for dumbbells are from \cite{Licht:2020odx}. The same color coding is used in the next figures. \label{fig:JOmegaDiagram} 
 }
 \end{figure}

Fig.~\ref{fig:JOmegaDiagram} is a phase diagram that summarizes the main properties of these solutions, and the implications for the possible initial and, more importantly, final states of a collision. Depending on the value of $J/M$ (distinguished by band-colouring in the figure) the stable phases of stationary single blobs are:
{\renewcommand{\arraystretch}{1.3}
\begin{center}
\begin{tabular}{rl}
$0 \leq J/M < 2$:& MP black holes\\
$2 \leq J/M < \frac{4}{\sqrt{3}}\approx 2.31$:& Black bars \\
$\frac{4}{\sqrt{3}} \leq J/M <  (J/M)_c$:& Black dumbbells \\
$(J/M)_c \leq  J/M$:& No stable single black hole 
\end{tabular}
\end{center}}
\noindent where the numerically determined upper limit $(J/M)_c$ for the existence of stable phases is \eqref{JMc}.

\begin{figure}[h!]
\centering
\includegraphics[width=0.6 \linewidth]{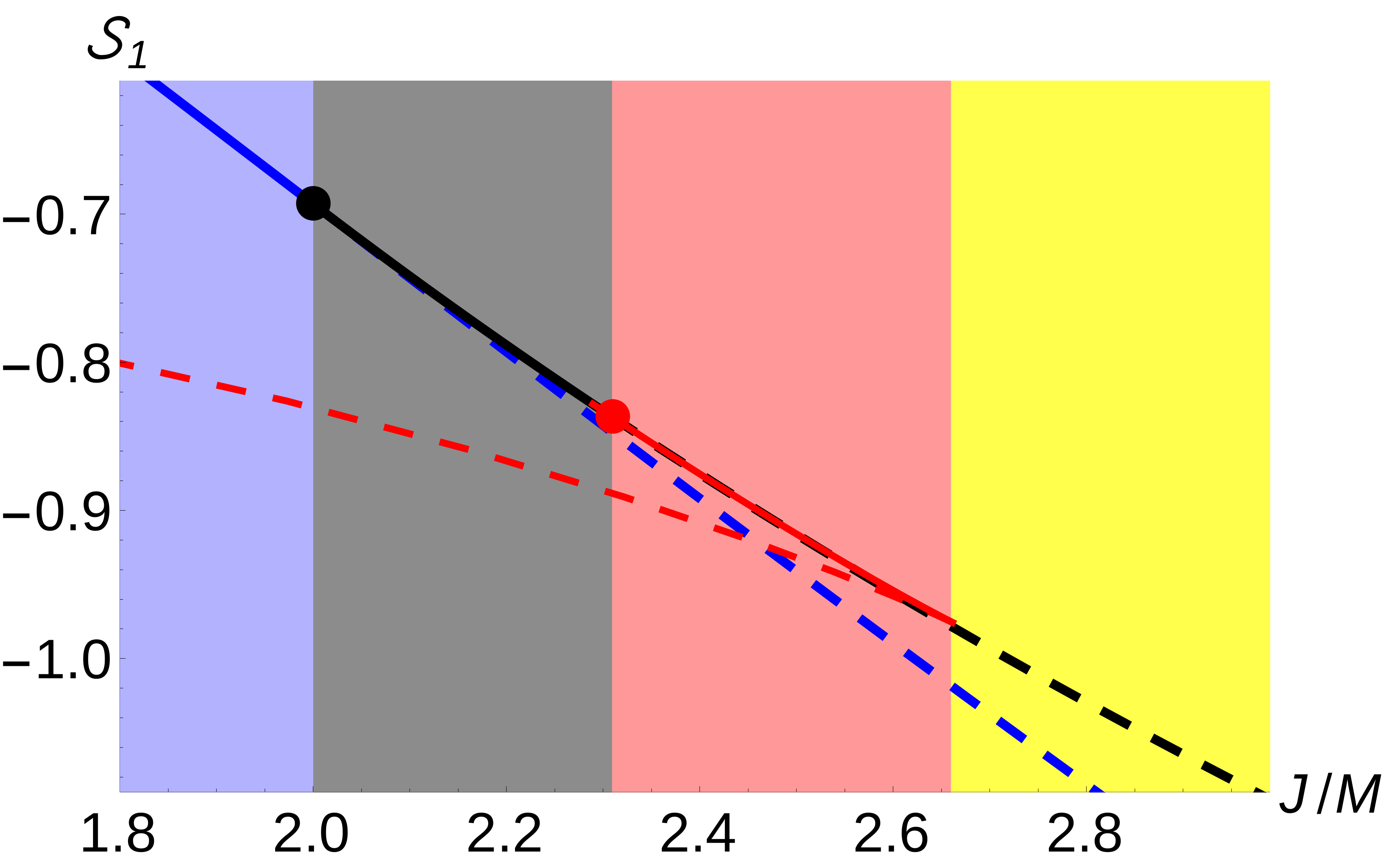}
\caption{\small Phase diagram depicting the entropy $\mathcal{S}_1$ of the different configurations close to the first black bar zero-mode, as a function of angular momentum. For $J/M> 2$ the black bar (black) is entropically favored over the MP black hole (blue). At the zero-mode $J/M = 4/\sqrt{3}$, a branch of stable dumbbells (red) appears with $\mathcal{S}_\text{inv}$ slightly higher that that of unperturbed black bars. This phase dominates entropically up to the turning point at $(J/M)_c\approx 2.66$, where stable dumbbells cease to exist, and the system typically evolves to a fission. \label{fig:JEntropyDiagram}}
\end{figure}

Let us clarify an aspect of the stability of phases in this diagram that was not discussed in \cite{Licht:2020odx}. In that article a second branch of dumbbells (lower in $\Omega$, shown dashed in fig.~\ref{fig:JOmegaDiagram}) was found to exist, starting from $J/M=0$ until it joins the first, upper branch at $(J/M)_c$. In this second branch, the dumbbells are more like slowly rotating black hole binaries, consisting of two gaussian blobs joined by a thin, long tube between them. All these solutions have the same LO entropy, but the NLO entropy $\mc{S}_1$ \eqref{calS1} distinguishes between them.
In fig.~\ref{fig:JEntropyDiagram} we show that lower-branch dumbbells have less entropy than the upper branch. They are therefore thermodynamically unstable. Moreover, a Poincar\'e turning point argument tells us they must have one more negative mode than the upper-branch, and hence be dynamically unstable. This is indeed consistent with two other observations: (i) in our numerical collisions, we never observe a lower-branch dumbbell forming (while upper-branch dumbbells do form); (ii) stationary Keplerian binaries in $D\geq 6$ exist but are unstable.

In contrast, upper-branch dumbbells resemble (segments of) stable non-uniform black strings (fig.~\ref{fig:critdumb}). Although other more non-uniform phases were found in \cite{Licht:2020odx}, by generic turning-point/bifurcation arguments they are expected to have more negative modes and hence be dynamically unstable. Therefore, no other stable solutions are expected to exist besides those shown in fig.~\ref{fig:JOmegaDiagram}.

Fig.~\ref{fig:JOmegaDiagram} is then a major guide to predicting the outcome of a collision between two blobs, based only on the total angular momentum $J$ and total mass $M$ of the system, which are conserved. If $J/M<2.66$, two blobs that merge can relax into a stable single blob, namely the only one that is stable for the corresponding value of $J/M$. Bear in mind that they will not necessarily do so, since the dynamical evolution may avoid that endpoint.

If $J/M>(J/M)_c$ the final state must consist of more than a single blob. That is, if there is fusion it will be followed by fission.
If $J/M$ is less than $\approx 4$, we observe that the end state is always two outgoing black holes, \ie $2\to 1\to 2$. At higher $J/M$, a third, smaller black hole can appear between them, the apparent reason being that at these angular momenta there exists an unstable branch of three-bumped bars. End states with more than two blobs are entropically disfavored but nevertheless are dynamically possible. In particular, in collisions at large $J$ with large initial orbital angular momentum the evolution can exhibit complicated patterns. Their investigation would take us beyond the scope of this article.

\subsection{$2\to 1$ vs.\ $2\to 1\to N$}\label{subsec:2to}


In the next sections we will study collisions between two initial MP black holes, of equal mass and equal initial spins, with initial rotation parameter $a_{\text{in}}$ within the stability range of MP black holes $0\leq a_{\text{in}}<1$.\footnote{We do not consider initial black bars and dumbbells. They are expected to exist at finite $D\geq 6$ but not be completely stable, not even stationary, since they must radiate gravitational waves.} Their initial velocities  will be $\pm u_{\text{in}}$ and the impact parameter $b_{\text{in}}$. We select the collisions where there is fusion; the cases where the two black holes scatter without ever merging may also be of interest but their physics is different than we intend to explore here and we will barely discuss them\footnote{Since the effective theory describes a continuous horizon, the distinction is not perfectly clear-cut and involves discretionary choices. However, the more ambiguous cases are only marginal to our analysis.}.

For some values of the initial parameters $(a_{\text{in}}, u_{\text{in}}, b_{\text{in}})$ the system fissions. 
\begin{figure}[t]
\centering
\includegraphics[width=0.4 \linewidth]{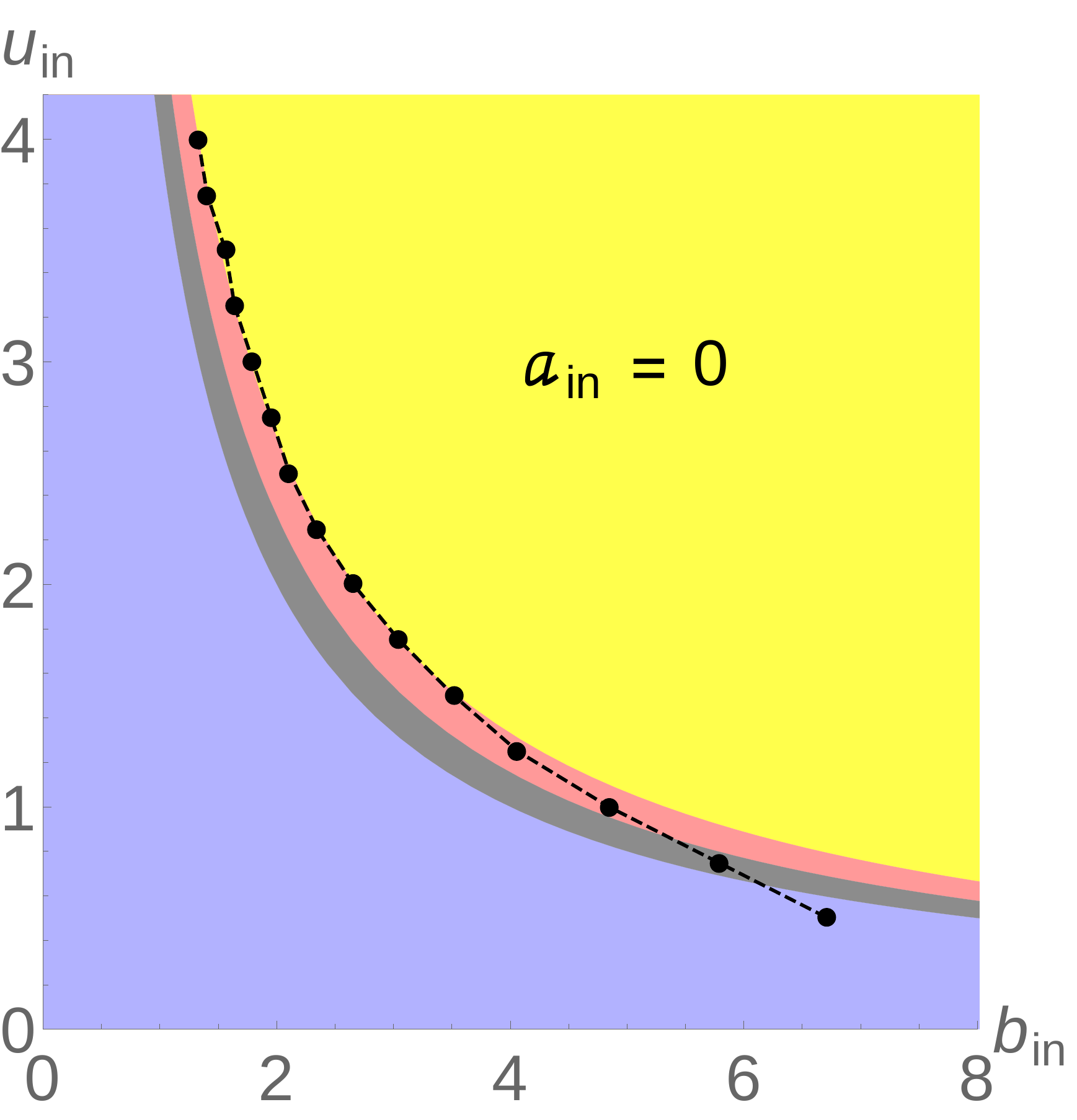}
\qquad
\includegraphics[width=0.4 \linewidth]{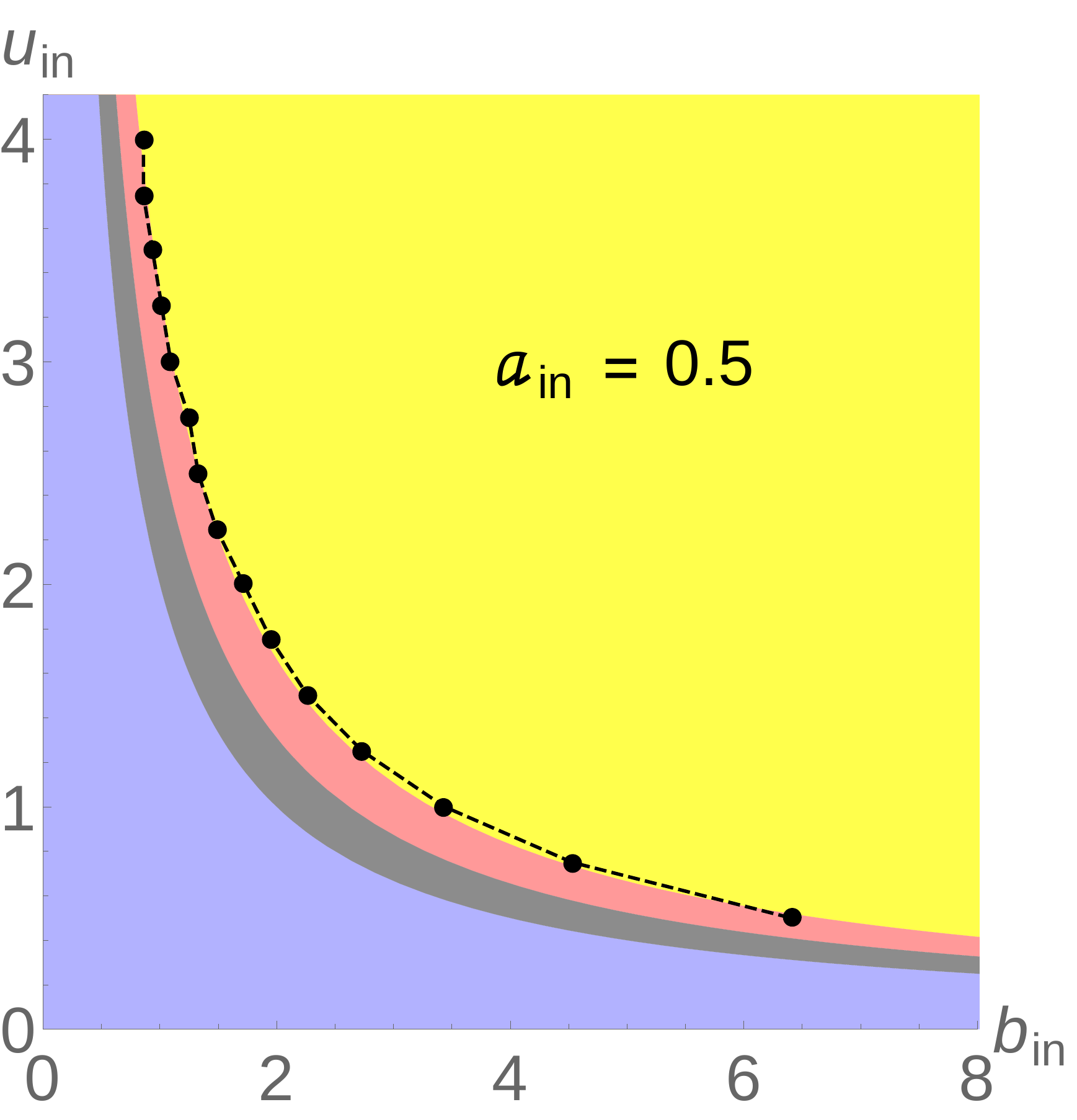}
\caption{\small Outcome of symmetric collisions of two black holes with initial spin, velocity, and impact parameter  $(a_{\text{in}}, u_{\text{in}}, b_{\text{in}})$. The dots (joined by dashed lines) separate between initial conditions that lead to $2\to 1$ fusion events (below the dots) and $2\to 1\to N$ cosmic-censorship-violating fission events (above the dots). The colors distinguish between the stable phases available (same color coding as in fig.~\ref{fig:JOmegaDiagram}). For small enough $b_{\text{in}}$ the system always settles down into the available stable single blob, but for very large $b_{\text{in}}$ the dynamical evolution passes too far from the stable phase and proceeds to fission. \label{fig:CCVBoundary}}
\end{figure}
Figure \ref{fig:CCVBoundary} shows the numerically determined boundary (dots joined by dashed lines) between initial conditions that lead to a $2\to 1$ collision, leaving a rotating central object, and initial conditions that produce more than one outgoing object, usually two but possibly more.

As already noted in previous papers  \cite{Andrade:2018yqu,Andrade:2019edf}, the boundary follows a curve  of constant $J/M = (J/M)_c$, as long as the impact parameter is below some threshold (which depends on $a_{\text{in}}$). This means that for these values of $b_{\text{in}}$ the merger always settles down into the unique stable solution that is available. However, if $b_{\text{in}}$ is large enough, this possible end state is avoided: the two colliding black holes form a horizon that is too elongated to find its path to the stable blob. In these cases, the colliding black holes attract each other deflecting their trajectories,\footnote{Despite the absence of stable Keplerian orbits in $D\geq 5$, we find that the two blobs can perform more than one revolution around each other before either flying apart or merging. In this respect, these collisions resemble four-dimensional ones more than one might have expected.} and then they either fly apart or suddenly fall onto each other to form a stable central object. It is suggestive that this $2\to 2$ scattering might be understood as the formation of an unstable, long dumbbell (two gaussian blobs joined by a long tube), which then either collapses or breaks apart.


\subsection{Kinematics, entropy and geometry of the collisions}\label{subsec:kinent}

Let us now be more specific about the collisions we study.
The two initial black holes are MP blobs on a 2-brane like \eqref{MPblob} that start in the $(x,y)$ plane at 
\beq
(x,y)=\lp\pm \infty,\pm \frac{b_{\text{in}}}2\rp\,,
\eeq
with velocities 
\beq
(v_x,v_y)=(\mp u_{\text{in}},0)\,.
\eeq
The latter are achieved by applying a Galilean boost to each blob. The entropy of each individual boosted black hole, normalized by its own mass, is
\beq\label{S1MP}
\mc{S}_1=-\frac12 u^2  -\ln (1+a^2)
\eeq
(see eq.~\eqref{S1boost}). The presence of the term $\propto u^2$ is due to the fact, already mentioned, that the kinetic energy must be subtracted from the total energy of the black hole, since only the rest (irreducible) mass contributes to entropy. It is directly related to the fact that the horizon area of a black hole does not change through Lorentz contraction \cite{Horowitz:1997fr}, which is also a property of entropy in the effective theory, as proved in appendix~\ref{app:boost}. 

The entropy of a system of two equal MP black holes, now normalized by their combined mass $M$, is
\beq\label{2bhS1}
\mc{S}_1=-\frac12 u^2 -\ln 2(1+a^2)\,.
\eeq
and, using \eqref{JMa}, their total angular momentum is\footnote{Each black hole has spin $2(M/2)a$ and orbital angular momentum $(M/2) u (b/2)$.}
\beq\label{angmom2bh}
\frac{J}{M}=2 a+\frac{b u}2\,.
\eeq

The conservation 
of mass-energy (which includes kinetic energy to NLO) and angular momentum imposes restrictions on the possible outgoing final states, and on how much entropy can be produced. 
The analysis can be made entirely within the large $D$ effective theory, but since we are only considering initial and final states that are MP black holes, we could also consider the  properties of the known solutions exactly in $D$. That is, we could work at finite $D$ using physical magnitudes, expand in $1/D$ and translate into effective theory magnitudes. The two methods of calculation are easily seen to agree.\footnote{Similar considerations were made in \cite{Emparan:2003sy}.}

\paragraph{$2\to 1$: fusion.} If the black holes fuse and then relax into a single, stable blob, this final state can be read from fig.~\ref{fig:JOmegaDiagram} as the unique stable solution with the value
\beq\label{angmomfus}
\frac{J}{M}=2 a_{\text{in}}+\frac{b_{\text{in}} u_{\text{in}}}2\,.
\eeq
If this end state is an MP black hole or a black bar, the final entropy will be
\beqa
\mc{S}_1^{\rm(MP)}&=&-\ln\lp 1+\frac14 \lp\frac{J}{M}\rp^2\rp^2\,,\\
\mc{S}_1^{\rm(bar)}&=&-\ln\lp\frac{J}{M}\rp\,.
\eeqa
The total entropy production in the fusion will be the difference between these and \eqref{2bhS1}.
We do not have analytical expressions for the entropy of black dumbbells.

\paragraph{$2\to 1\to 2$: fusion $\Rightarrow$ fission.} The final states of the $2\to 1\to 2$ collision are always two equal MP black holes\footnote{In principle, these could also be stable bars and dumbbells. However, the spin of the outgoing states that we observe is always below the range of their existence.} with outgoing parameters $0\leq a_{\text{out}}<1$ and
\beq\label{xyout}
(x,y)=R(\theta)\lp\pm \infty,\pm \frac{b_{\text{out}}}2\rp\,,
\eeq
\beq\label{vout}
(v_x,v_y)=R(\theta)(\pm u_{\text{out}},0)\,,
\eeq
where $R(\theta)$ is a rotation matrix with angle $\theta$.
Conservation of mass and angular momentum implies
\beq\label{angmomcons}
\frac{J}{M}=2 a_{\text{in}}+\frac{b_{\text{in}} u_{\text{in}}}2=2 a_{\text{out}}+ \frac{b_{\text{out}} u_{\text{out}}}2\,.
\eeq

The collision is characterized by seven parameters: $(a,u,b)_\text{in/out}$ plus the scattering angle $\theta$. The latter is not affected by conservation laws and it does not enter into entropic arguments, so we will leave it aside in the following discussion.

Of the six parameters $(a,u,b)_\text{in/out}$, only five are independent once \eqref{angmomcons} is imposed. We can regard the three initial parameters as given, and then two outgoing parameters, say, $u_{\text{out}}$ and $a_{\text{out}}$, are unconstrained by conservation laws, that is they will be determined by the dynamical evolution of the system.

The difference in the entropy between the initial and final states is
\beq\label{deltas}
\Delta \mc{S}_1=\frac{u_{\text{in}}^2-u_{\text{out}}^2}{2}+\ln \frac{1+a_{\text{in}}^2}{1+a_{\text{out}}^2}\,.
\eeq

The entropy of the final state will be larger if the outgoing velocities and spins are as small as possible, since both $a$ and $u$ tend to reduce the entropy of a MP black hole with fixed mass. However, they cannot be made arbitrarily small. The total angular momentum must be conserved, and even though \eqref{angmomcons} seems to allow for two unconstrained outgoing parameters, we cannot expect to have $a_{\text{out}},u_{\text{out}}\to 0$. For any $J\neq 0$ this would require that the outgoing impact parameter diverges, $b_{\text{out}}\to \infty$, which is unreasonable: if the two initial black holes do indeed collide and merge, the outgoing impact parameter will be comparable to the size of an intermediate (unstable) state with the same $J$ and $M$. From the distance between the two peaks in the critical dumbbell, fig.~\ref{fig:critdumb}, we can expect that
\beq\label{bdumb}
b_\mathrm{out}\approx 7\,,
\eeq
and probably a little larger after the fission.
This is indeed a good predictor for the actual values we find below. Another well-motivated and better defined geometric estimate is obtained by demanding that $b_\mathrm{out}$ is approximately equal to twice the radius of the two outgoing black holes. In appendix~\ref{app:bout} we find that this gives
\beq\label{boutguess}
b_\mathrm{out}\approx 2\sqrt{2(1+a_\mathrm{out}^2)\ln\epsilon_b^{-1}}\,.
\eeq
It depends on a small number that we estimate to be $\epsilon_b\approx 10^{-3}$. If entropy is to be maximized, then $b$ will be close to this upper bound.

Eqs.~\eqref{angmomcons}  and \eqref{boutguess} leave one unconstrained degree of freedom: an equation between $u_{\text{out}}$ and $a_{\text{out}}$.  The last constraint that fixes them will be discussed in sec.~\ref{sec:bhscat}.

\section{Entropy production}\label{sec:entpro}

With our methods we can easily track the entropy production, in space and in time, during the evolution of three different kinds of phenomena:
\begin{itemize}
\item $1\to N$: decay and fission of unstable black holes
\item $2\to 1$: fusion of two black holes
\item $2\to 1\to 2$: fusion of two black holes followed by fission
\end{itemize}
Understanding entropy production in the first two will give us insight into the third.

We evolve the equations numerically, using two different codes (the same as in  \cite{Andrade:2018yqu,Andrade:2019edf}, now using finite differences instead of FFT differentiation), until the system either settles into a stable single blob, or breaks up into blobs that fly apart. By keeping track of $m(t,\mathbf{x})$ and $v^i(t,\mathbf{x})$ we can then compute all physical magnitudes.



\subsection{$1\to N$: decay and fission of unstable black strings and black holes} 

\begin{figure}[h!]
\centering
\includegraphics[width=.95\linewidth]{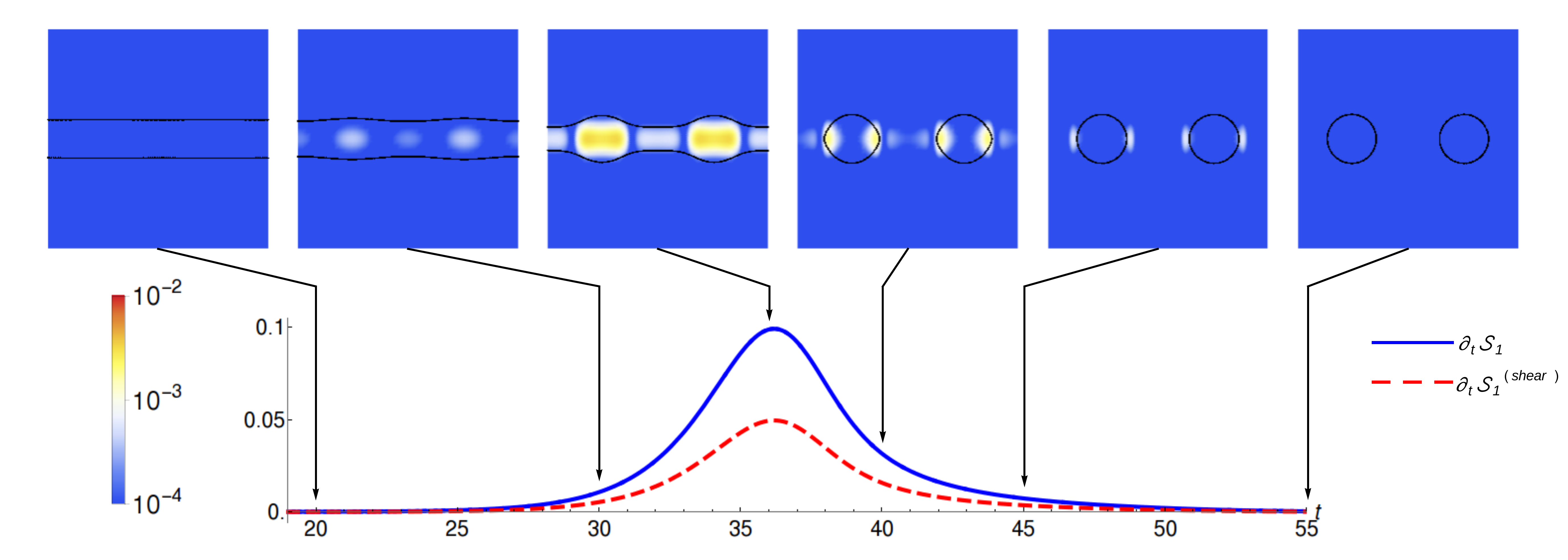}
\\
\medskip
\includegraphics[width=.95\linewidth]{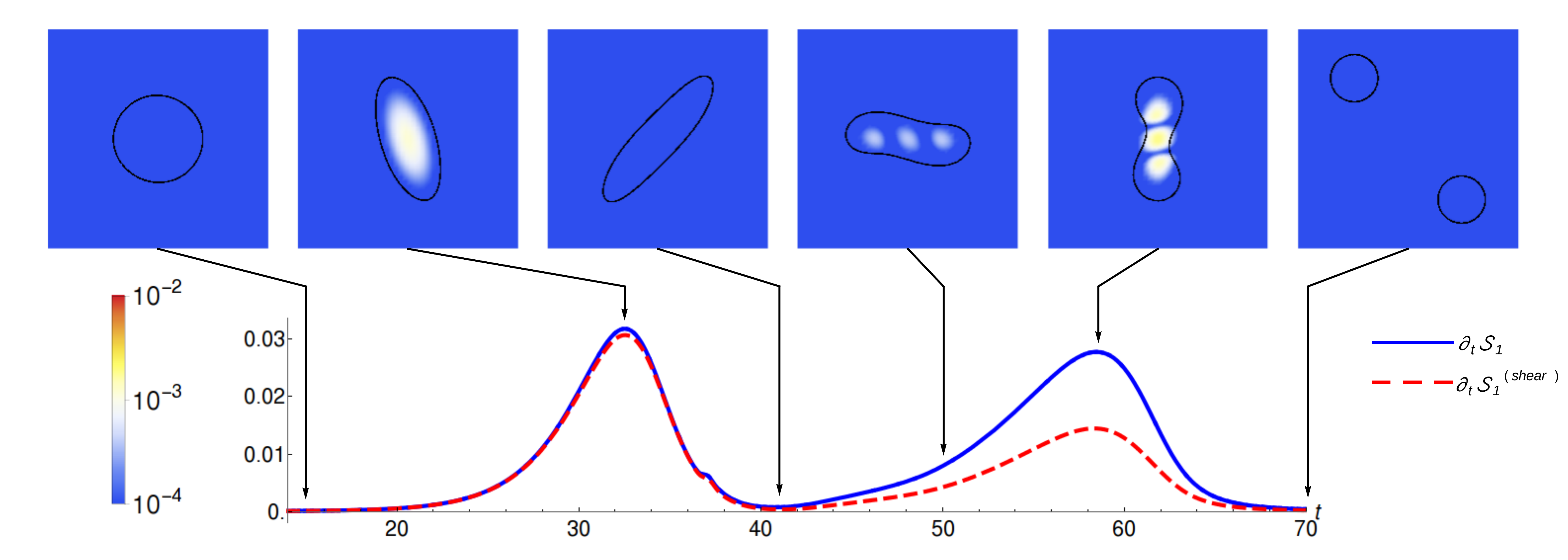}
\\
\medskip
\includegraphics[width=.95\linewidth]{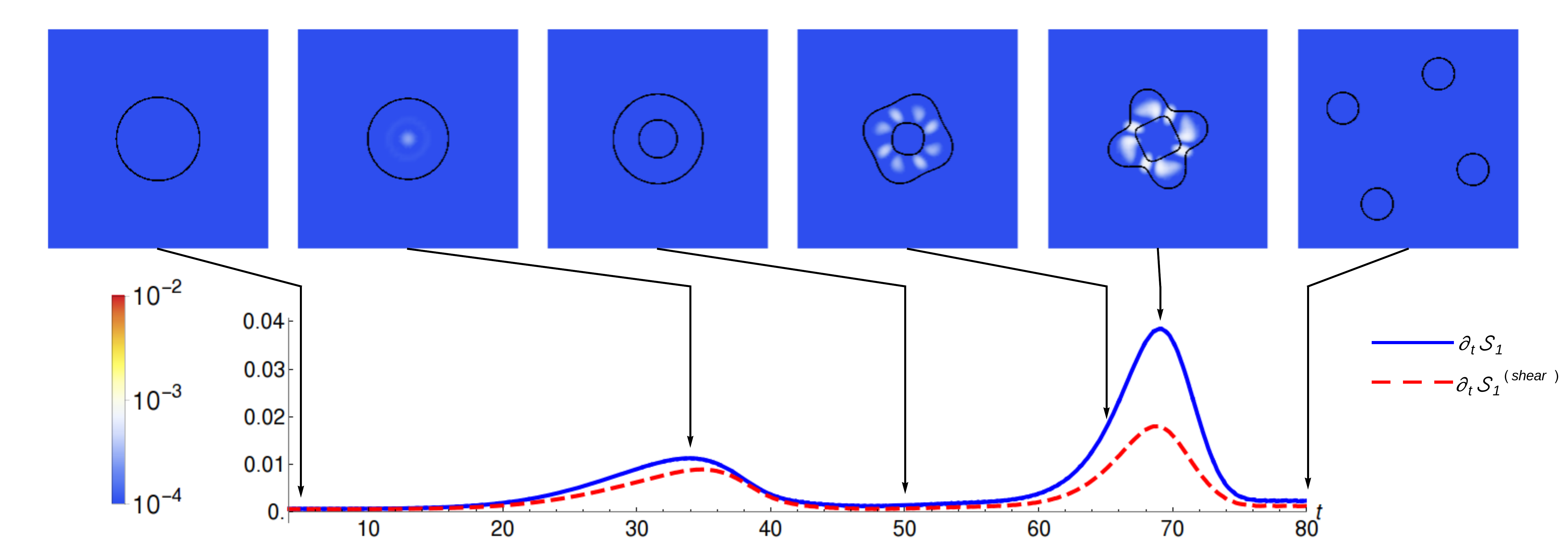}
\medskip
\caption{\small Entropy production, as a function of time and space, during the decay of: unstable black string (top); ultraspinning MP black hole through intermediate black bar (middle); ultraspinning MP black hole through intermediate black ring (bottom). The blue curves give the production rate of the NLO entropy, $\partial_t \mc{S}_1$ (integrated in space); the dashed red curves are the production rate through dissipation of shear (and not of expansion), $\partial_t \mc{S}_1^{\rm{(shear)}}$. The density plots show the time derivative of the entropy density (colors in log scale). The thin black  contours serve to guide the eye to where the black hole blobs are, and correspond to $m(x,y) = 0.001 M$. We can see that the pattern of entropy production in the black string decay is reproduced in the second peak of the decays of the MP black hole. The first peak is mostly due to shearing when the intermediate black bar or black ring forms.
\label{fig:decays}}
\end{figure}

Here we follow the non-linear evolution of the decay of an unstable, fissile blob. We have chosen three important examples which most clearly exhibit the physics relevant for other more complex evolutions, see fig.~\ref{fig:decays}: the black string; an ultraspinning MP black hole with $a=2$ decaying through an intermediate black bar; and an MP black hole with $a=3$ decaying through an intermediate black ring. The latter evolutions are triggered by choosing different inital perturbations, which excite different unstable modes of the ultraspinning black hole.

Let us emphasize that our simulations of the decay of unstable black strings are not expected to reproduce the details of the late-time evolution in the (much more expensive) numerical evolutions in \cite{Lehner:2010pn}, nor of the related and more complex simulations in \cite{Figueras:2015hkb,Figueras:2017zwa,Bantilan:2019bvf}. This has been discussed in detail in \cite{Emparan:2018bmi}. In particular, the large $D$ effective theory does not reveal the cascading formation of small `satellites'. However, our concern here is with how entropy is produced in this decay, and this appears to occur mostly in the intermediate stages of the evolution. At late times, most of the mass and area reside in the large, black-hole-like blobs, and little on the satellites and thin tubes inbetween them. Therefore, we expect that our study accounts for the main contributions to entropy production.

The analysis of black string decay reveals generic aspects of entropy production in fission. The pattern we see in fig.~\ref{fig:decays} (top) will be present in all subsequent fission phenomena: a single peak in the entropy production rate, midway along the fission, with dissipation equally shared into shear and expansion. 

The two decays of the ultraspinning MP black hole in  fig.~\ref{fig:decays} show qualitative similarities between themselves: first, a long-lived but ultimately unstable configuration forms---a black bar, or a black ring.\footnote{When $D$ is not large enough this bar radiates away its excess spin fast enough to return to stability \cite{Shibata:2010wz}.} 
Entropy is generated mostly through shear dissipation. In fact, it is clear that a dominant shearing motion must be driving the evolution to a bar, while the formation of the ring should also involve some compression. Both features are visible in the entropy production curves. Afterwards, this intermediate state decays following the pattern of black string fission.

The second peak in entropy production appears to have universal features. This confirms that the physics of black string decay also controls the fission of the blob. Observe, however, that in the MP decay the peak is a little higher---hence more irreversible---than in the black string. This could be expected since the latter is a more symmetric configuration.

Finally, we see that the duration of the string break up is on the order of
\beq\label{tfission}
(\Delta t)_\textrm{fission} \approx 20 M\,.
\eeq
This will be a characteristic of other fissions.


\subsection{$2\to 1$: fusion $\Rightarrow$ thermalization}

In a $2\to 1$ collision the final state is completely determined by the conserved initial value of $J/M$: the system settles into the only stable stationary black hole with that value: an MP black hole, a black bar, or a black dumbbell. 

\begin{figure}[t]
\centering
\includegraphics[width=1\linewidth]{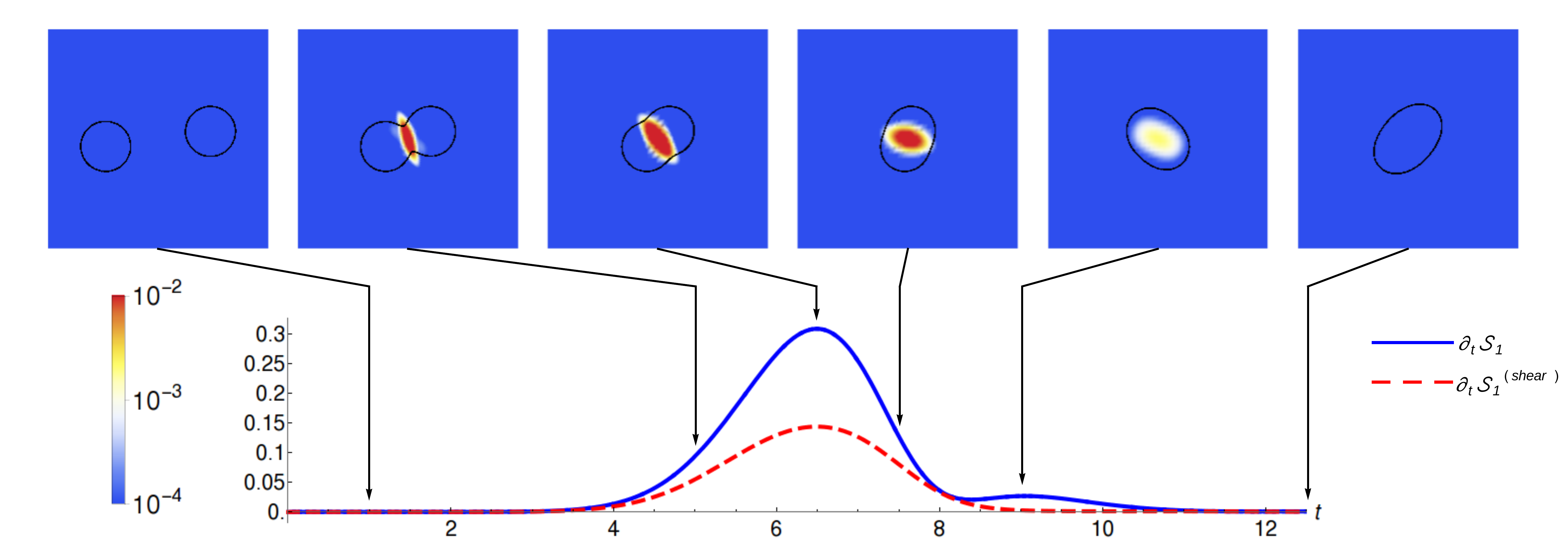}
\caption{\small Entropy production during a collision with fusion into a stable black bar. The first, large fusion peak is followed by a smaller peak for the thermalization to the black bar. The height of this second peak is comparable to that in fig.~\ref{fig:decays} (middle). \label{fig:fusion}}
\end{figure}

In fig.~\ref{fig:fusion} we present an illustrative example: a symmetric collision of two black holes with $(b_\text{in},u_\text{in},a_\text{in}) = (2, 1.0, 0.6)$, so  $J/M=2.2$, resulting in the formation of a stable black bar.
The figure shows that when the two black holes first meet and fuse there is a large production of entropy. There follows a phase in which the system equilibrates (thermalizes) into the final stable black bar. This second phase is similar to the formation of the (unstable) bar in the decay of the ultraspinning MP black hole in  fig.~\ref{fig:decays} (middle), with both peaks having similar height ($\approx 0.03$, in mass-normalized entropy rate). The duration of this phase in the decay of the MP black hole is much longer, since the system there starts in stationary, but unstable, equilibrium, while the merged horizon is farther from equilibrium. 

The contributions to dissipation from shear and expansion vary at different stages in the evolution. During the initial fusion phase, one or the other may dominate depending on the initial parameters, but generically we observe both expansion (and compression) and shearing motion of the blob, which contribute roughly equally to entropy production. During the formation of the intermediate quasi-thermalized bar, the proportions of shear and expansion can vary, depending on how much the preceding blob is already bar-like or not. In the simulation shown in fig.~\ref{fig:fusion}, from $t\approx 9$ to $1$ the blob has to undergo less shearing to acquire the bar shape than in fig.~\ref{fig:decays} (middle) from $t\approx 20$ to $40$. Presumably this explains the lower presence of shear dissipation.

\subsection{$2\to 1\to 2$: fusion $\Rightarrow$ quasi-thermalization $\Rightarrow$ fission}

With large enough total angular momentum, the fusion results in a fissile intermediate state.  

\paragraph{Stages in the evolution.}

\begin{figure}[t]
\centering
\includegraphics[width=1\linewidth]{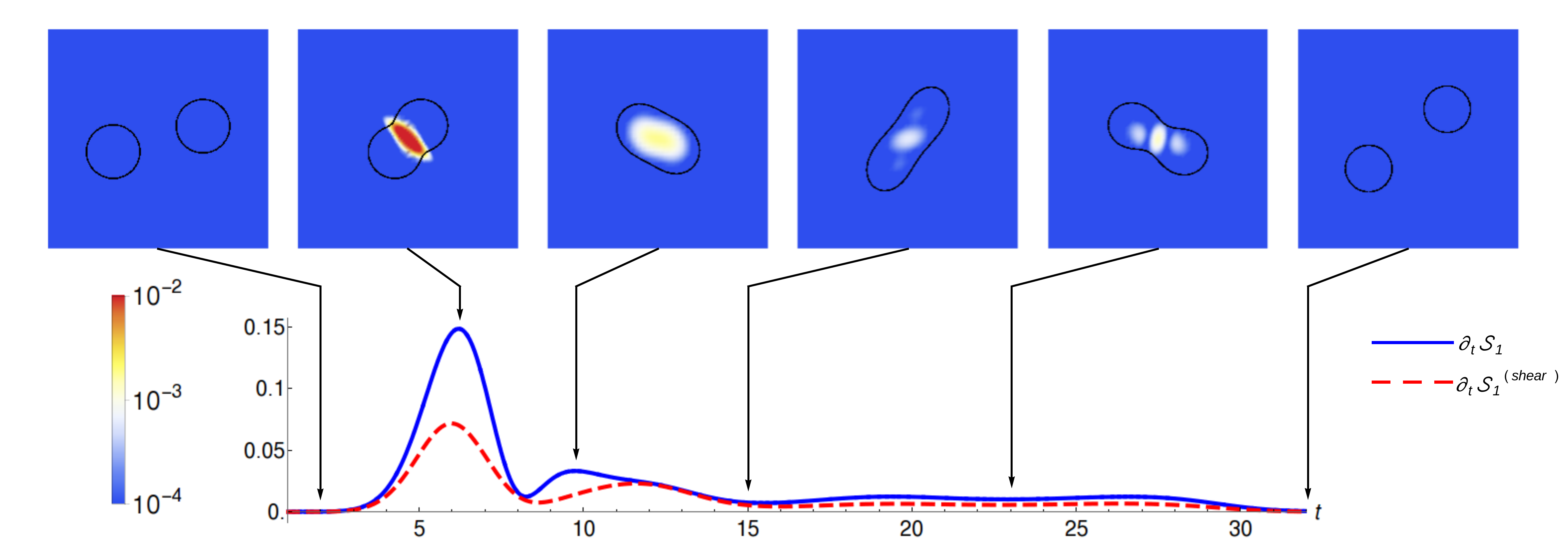}
\caption{\small Entropy production during a collision with fusion followed by fission. The entire process can be divided in three stages: $t\approx 4$ to $8$, fusion; $t\approx 9$ to $14$, quasi-thermalization into a bar; $t\approx 15$ to $30$, fission of the bar, similar to black string decay. \label{fig:EntropyFrames}}
\end{figure}

In fig.~\ref{fig:EntropyFrames} we present the evolution of entropy production in a collision with initial parameters $(b_\text{in},u_\text{in},a_\text{in}) = (3.4, 1.0, 0.8)$, so $J/M=3.3$. It can be interpreted by combining what we have learned so far:

\begin{enumerate}
\item \emph{Fusion:} $t\approx 4$ to $8$. This is a strongly irreversible phase, very similar to the first peak in $2\to 1$ fusion, fig.~\ref{fig:fusion}. 

\item \emph{Quasi-thermalization:}  $t\approx 9$ to $14$. The fused blob follows qualitatively the evolution of unstable MP black holes in fig.~\ref{fig:decays}~(middle): a quasi-thermalization phase with the (faster) formation of a long-lived bar.

\item \emph{Fission:} $t\approx 15$ to $30$. The intermediate bar fissions into two outgoing black holes, in a manner similar to fig.~\ref{fig:decays}~(middle), ultimately patterned after the decay of black strings, fig.~\ref{fig:decays}~(top). It lasts for a time comparable to \eqref{tfission}.
\end{enumerate}

Observe that not only the qualitative features of the decay of the MP bhs are reproduced in the $2\to 1\to 2$ collision: also the height of the peaks in the mass-normalized entropy production rates, after the fusion peak in fig.~\ref{fig:EntropyFrames}, are quantitatively similar.

The proportions of viscous dissipation from shear and expansion also follow what we have seen before. 

\section{Scattering of black holes and entropic attractors}
\label{sec:bhscat}

We now turn to a more complete investigation of collisions, in particular those that result in fission, and the role that the entropy increase plays in them. For this purpose 
we have performed an extensive, although not exhaustive, study of symmetric collisions of black holes for wide ranges of the initial parameters $(a_{\text{in}}, u_{\text{in}}, b_{\text{in}})$.

In our simulations we verify that the final blobs can be identified with known stable stationary blobs. In $2\to 1$ and $2\to 1\to 2$ events, the final spin parameter, $a_{\text{out}}$, is extracted from the width of the gaussian blobs by linear regression of $\ln m$ as a function of $r^2$, where $r$ is the distance to the center of the blob (see \eqref{MPblob}). In fission, we extract the parameters 
 $u_{\text{out}}$ and $b_{\text{out}}$ of the outgoing blobs from the velocity field at their centers,


As we have seen, fusion into a single stable black hole is fully determined by the conservation of mass and angular momentum. In our simulations we have been able to verify that the integration over space and time of the entropy density reproduces correctly the exact predictions from sec.~\ref{subsec:kinent}. This is a good check on the accuracy of our methods.

\subsection{$2\to 1\to 2$: In \& Out}\label{subsec:innout}

In contrast to $2\to 1$ fusion, here there is a continuous two-dimensional range of out states that are allowed by the conservation laws.
\begin{figure}[t]
\centering
\includegraphics[width= \linewidth]{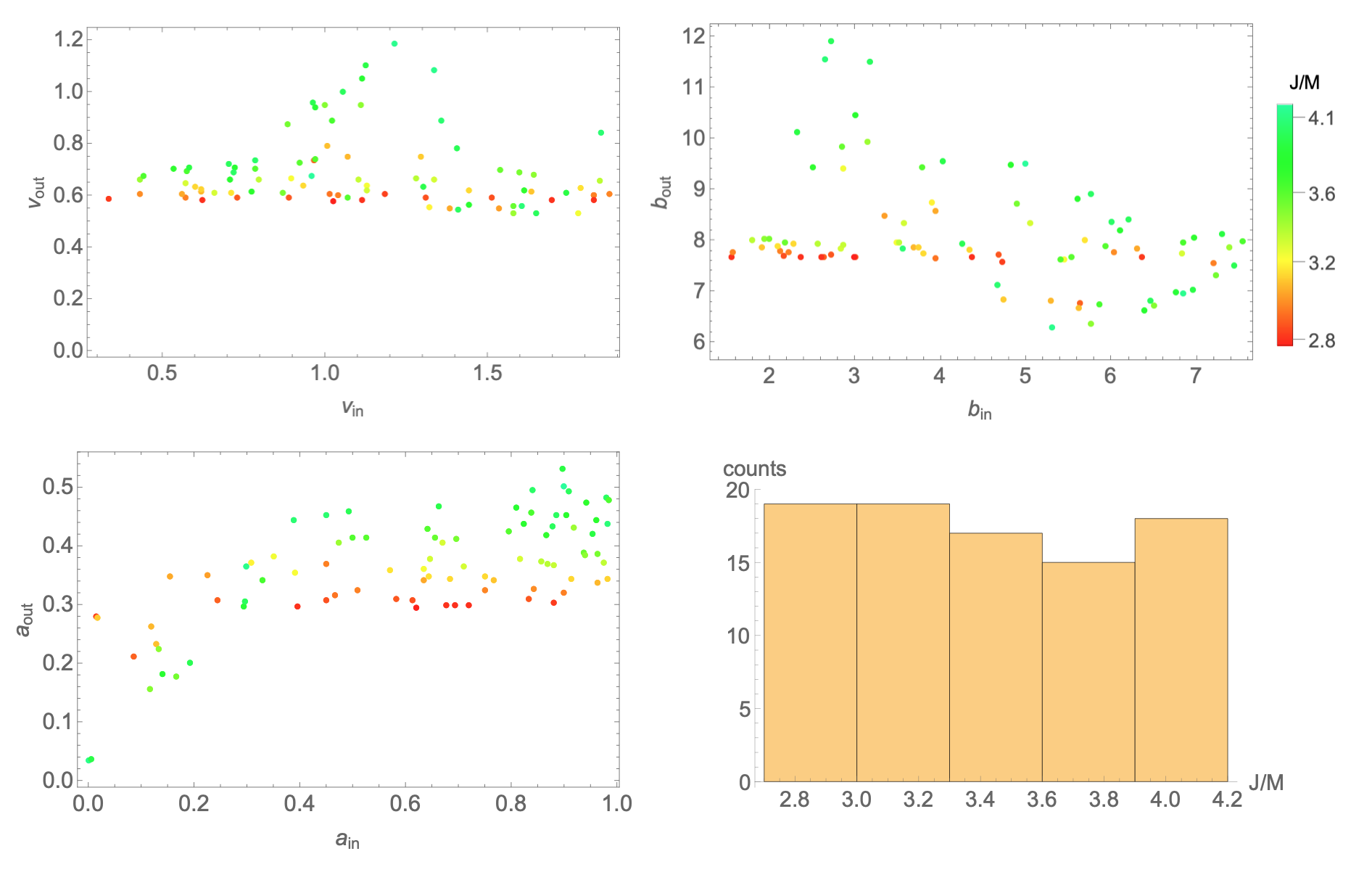}
\caption{\small $2\to 1\to 2$ collisions: in 100 simulations, the outgoing parameters $(a_{\text{out}}, u_{\text{out}}, b_{\text{out}})$ show little correlation with initial ingoing parameters  $(a_{\text{in}}, u_{\text{in}}, b_{\text{in}})$. This is a consequence of quasi-thermalization in an intermediate stage in the collision. Outgoing parameters cluster in a relatively narrow range, the more so the lower the total $J/M$, for which the intermediate phase lasts longer.
\label{fig:innout}}
\end{figure}
In figs.~\ref{fig:innout} we present the results for the outgoing parameters $(a_{\text{out}}, u_{\text{out}}, b_{\text{out}})$ of 100 simulations with randomly chosen values of the ingoing parameters  $(a_{\text{in}}, u_{\text{in}}, b_{\text{in}})$.\footnote{We discard events where the intermediate interaction is too weak to involve fusion. We implement this by removing from our analysis events for which the initial impact parameter is large ($b_{\text{in}} >7$)
and the outgoing parameters change less than $5\%$ relative to the initial ones. In these cases the spin $a$ changes very little, while $u$ and $b$ can change more, as expected if the process is one of direct (fusionless) $2\to 2$ scattering.} We also present, as color shading, the value of the the conserved $J/M$ for each event. Since our sampling is not exhaustive, we have not attempted to perform detailed statistical analyses, but nevertheless there are several discernible patterns in these plots that are worth remarking on.

First, there is a clear clustering of the out parameters. It is stronger the lower $J/M$ is, with $(u_{\text{out}}, a_{\text{out}}, b_{\text{out}})$ being essentially unique for the lowest $J/M$ (slightly above $(J/M)_c=2.66$). The latter are the cases where an unstable but very long-lived intermediate state forms in the collision. The dissipation that happens in this intermediate phase effectively erases the memory of the initial state parameters, other than the conserved $J/M$. As $J/M$ grows larger, the intermediate state is less long-lived and less precisely defined, and the system retains more memory of the initial configuration (for instance, there is some correlation between the values of $a_{\text{in}}$ and $a_{\text{out}}$), resulting in more dispersion in the plots.

More generally, the plots show that the out parameters lie approximately in the following ranges:\footnote{These central values and variances are indicative and should not be taken literally; the dispersion is strongly correlated with $J/M$, and it is very low near the critical value.}

Spin:
\beq\label{arange}
a_{\text{out}} \approx 0.3\,\substack{+0.2 \\ -0.1}\,, 
\eeq
with the upper bound being fairly robust.

Velocity:
\beq\label{vrange}
u_{\text{out}}\approx 0.6\,\substack{+0.4 \\ -0.1}\,,
\eeq
with a strong bias towards the lower value, which is never below $\approx 0.5$.

Impact parameter:
\beq\label{brange}
b_{\text{out}}\approx 8\,\substack{+2 \\ -1}\,.
\eeq

The scant correlation of these results with the initial values other than the conserved $J/M$ is indicative of intermediate thermalization. 

The clustering values might possibly be compared with those in the decay states of unstable blobs in the yellow regions close to $(J/M)_c$ in fig.~\ref{fig:JOmegaDiagram}. An indication in this direction is \eqref{bdumb}, but we have not attempted to go further since this requires additional extensive numerical studies. 


Regarding the scattering angle in the final states \eqref{xyout}, \eqref{vout}, we have observed that $\theta$ can be robustly obtained from the numerical simulations. In particular, it is independent of the scheme implementing a regulator at small $m$, and of the values of the regulator as this is decreased. We expect, therefore, that in collisions at finite $D$ this scattering angle can also be obtained from the classical evolution before the naked singularity forms. However, this angle does not play any role in our study in this paper.

\subsection{Entropy increase}\label{subsec:totent}

\begin{figure}[t!]
\centering
\includegraphics[width=0.9\linewidth]{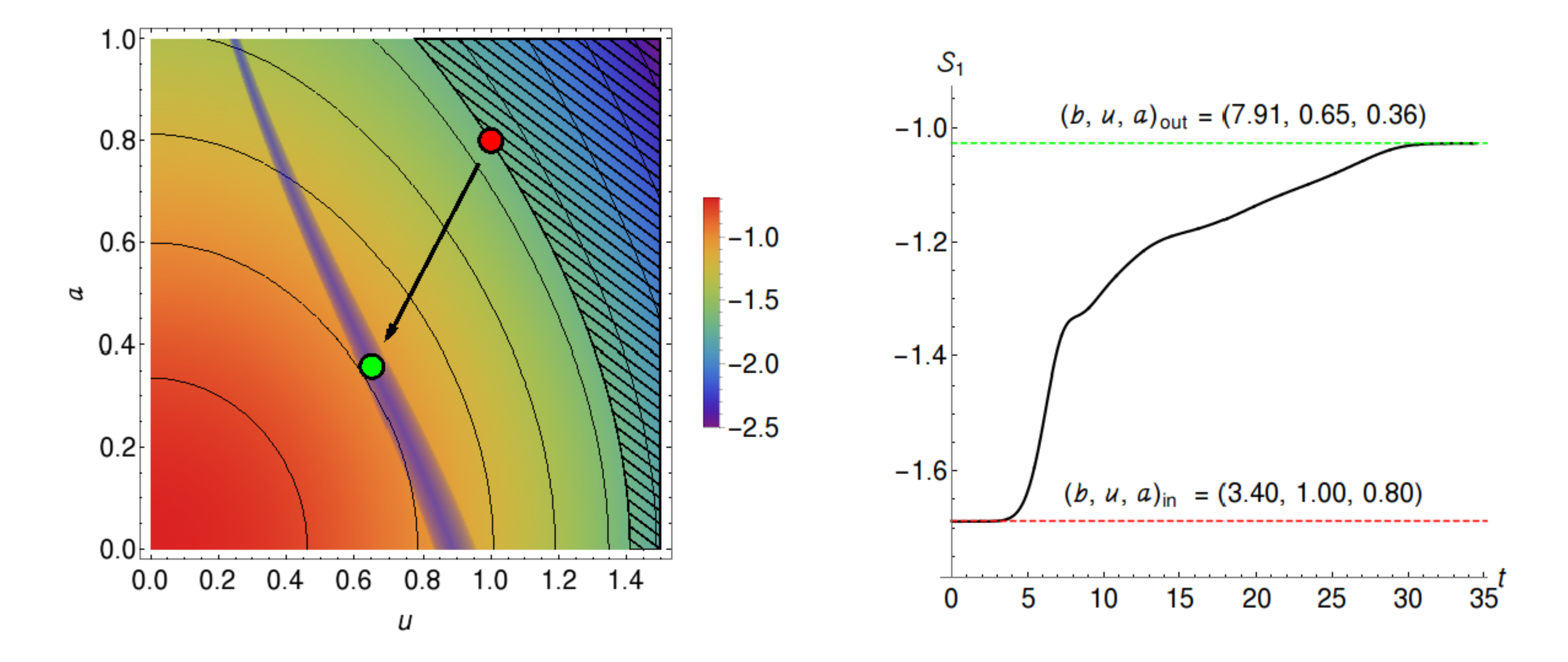}
\includegraphics[width=0.9\linewidth]{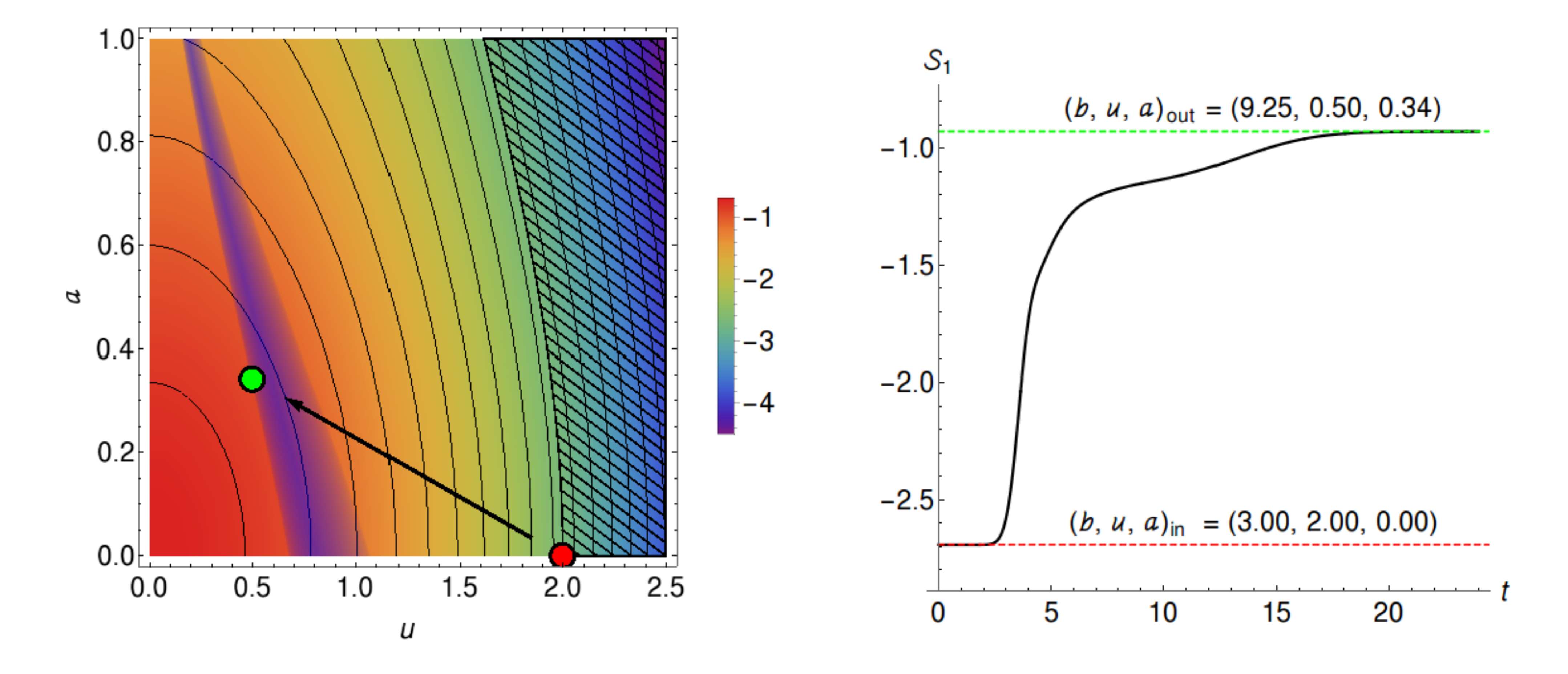}
\caption{\small Total entropy growth in symmetric $2\to 1\to 2$ collisions. \emph{Left}: coloured-contour plot for the total entropy of configurations with velocity and spin $(u,a)$, with a given value of $J/M$. Red and green dots indicate the initial and actual final states in the dynamical evolution. The hashed part marks final states forbidden by the second law. The purple band are states with impact parameter in the geometric range \eqref{boutguess}. We see that the actual final states lie very close to (although not exactly at) the maximum entropy with $b$ in this range. In particular, larger values of $a_\text{out}$ and smaller of $u_\text{out}$ would be entropically very disfavoured. \emph{Right}: evolution in time of $\mc{S}_1$ along the simulation. The initial and final parameters are (top)  $(b,u,a)_\text{in} = (3.4, 1.0, 0.8)$, $(b,u,a)_\text{out} = (7.9, 0.65, 0.36)$ and (bottom) $(b,u,a)_\text{in} = (3.0, 2.0, 0.0)$,  $(b,u,a)_\text{out} = (9.25, 0.5, 0.34)$. \label{fig:NeutralCollisionEntropy}}
\end{figure}

Recall that the initial and final states are characterized by the values of $(a,u,b)$,  one of which can be traded for the value of $J$, which is common for the initial and final states (we always set the total mass $M=1$). We find convenient to eliminate $b$, so in fig.~\ref{fig:NeutralCollisionEntropy}, for a given value of $J$, we  represent the initial and final states each one as a point (red and green, respectively) in the plane $(u,a)$. On this plane, we also show colour contours for the value of the mass-normalized entropy $\mc{S}_1$. We exclude the region of ultraspins $a>1$ since these MP black holes are unstable.

We already mentioned that the entropy cannot be fully maximized, since this happens at $(u,a,b)=(0,0,\infty)$. The geometric constraint \eqref{boutguess} (particularly good for low $J/M$) selects a set of possible final states, which we mark as a purple band in the $(u,a)$ plane.

In fig.~\ref{fig:NeutralCollisionEntropy} the entropy changes between initial and final states are shown in two illustrative cases.\footnote{The top right curve for $\mc{S}_1(t)$ is the integral of the blue curve in fig.~\ref{fig:EntropyFrames}.}  

The salient aspects of these plots are:
\begin{itemize}

\item Final states in the hashed region are excluded by the second law. High final velocities are then excluded. In particular, if the initial black holes are spinless, then the outgoing velocity cannot be higher than the ingoing one.

\item The entropy increases significantly, an in particular it is close to being maximized (but not fully maximized) among the possible outgoing states with geometrically-constrained impact parameter \eqref{boutguess}.\footnote{All purple bands are centered at $\epsilon_b = 0.001$, as defined in \eqref{meps}. The bands in figs.\ref{fig:attract} and \ref{fig:NeutralCollisionEntropy}~(up) span the range $\ln(\epsilon_b) = \ln(0.001) \pm 0.5$. In figs.~\ref{fig:NeutralCollisionEntropy}~(down) and \ref{fig:hiJattract} the range is wider, $\ln(\epsilon_b) = \ln(0.001) \pm 1.5$.}

\end{itemize}

In fig.~\ref{fig:NeutralCollisionEntropy}~(right) we  show the time evolution of the entropy. In the first one (top) the total entropy change is approximately equally subdivided between the sharp production at the beginning of the collision and the slower subsequent production rate. In the second one (bottom), which starts with very high initial velocity,  most of the entropy is quickly produced in the initial stages.

\subsection{Entropic attractors}

We can now combine the analyses of sec.~\ref{subsec:innout} and sec.~\ref{subsec:totent} to obtain a global perspective on the role of total entropy production in the evolution of the system.

\begin{figure}[t]
\centering
\includegraphics[width=0.6\linewidth]{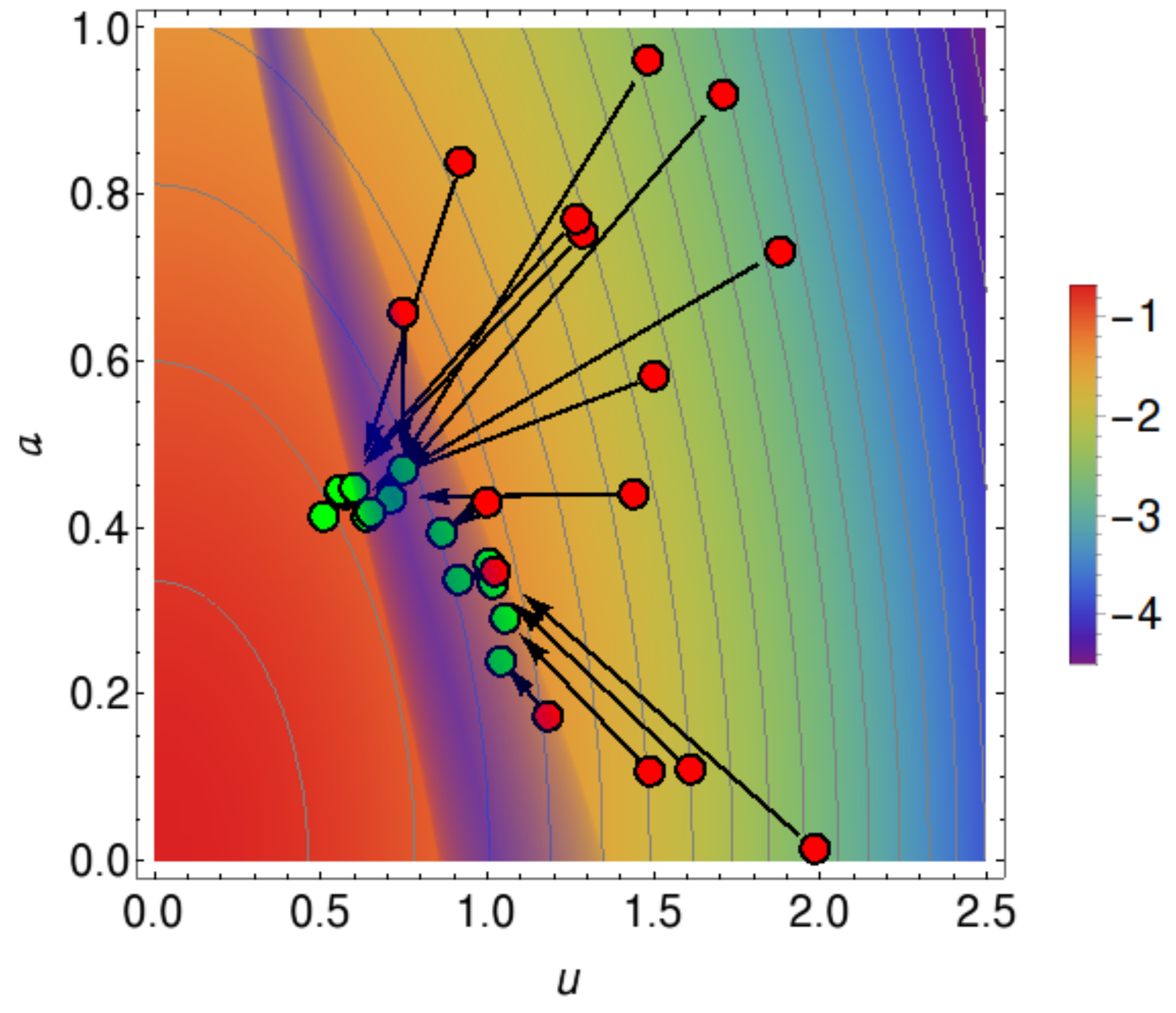}
\caption{\small Entropic attractor in collisions $2\to 1\to 2$  with $J/M=3.8$. See fig.~\ref{fig:attract} for the explanation. The attractor is less strong as $J/M$ grows larger. Note that initial states to the left of the purple band have large initial impact parameters and the black holes do not merge, so we do not include them.
\label{fig:hiJattract}}
\end{figure}

In figs.~\ref{fig:attract} and \ref{fig:hiJattract} we show the results of sampling a large number of collisions  $2\to 1\to 2$ with two specific values of $J/M$: a low value close to $(J/M)_c$ in fig.~\ref{fig:attract}, and a quite higher one in fig.~\ref{fig:hiJattract}. The clustering of the final states seen in sec.~\ref{subsec:innout} is even more clearly visible here, and also the near-maximization of the entropy that we discovered in sec.~\ref{subsec:totent}. The attractor that funnels the evolution is stronger the closer to the critical value of the conserved $J/M$, but fig.~\ref{fig:hiJattract} shows that it, and the near-maximization of the entropy, are present even when $J/M$ is quite far from criticality.

We conclude that the dynamical outcome of the collision can be approximately predicted, after imposing kinematic and geometric constraints, by near-maximization of entropy generation. The maximization is not exact, but this principle is a powerful guide to the end result of a  complex dynamical process.

\section{Charge diffusion in black holes}\label{sec:chadif}


Since the entropy of neutral black holes is proportional to their mass in the limit $D\to\infty$, it can only be generated at NLO in the $1/D$ expansion---although, as we have argued, this production can be computed using the LO effective theory.
However, when charge is present, the entropy of a black hole when $D\to\infty$ is no longer proportional to the mass. Instead of \eqref{bhentD}, we have
\beqa\label{SMQ}
S(M,Q)&\propto& \lp M+M\sqrt{1-2\lp\frac{Q}{M}\rp^2}\rp^{\frac{D-2}{D-3}}\nn\\
&=& M+M\sqrt{1-2\lp\frac{Q}{M}\rp^2}+\ord{\frac1{D}}\,.
\eeqa
This is easily seen to imply that in the fusion between two black holes with different charge-to-mass ratios, $Q_1/M_1\neq Q_2/M_2$ (including the charge sign), the charge redistribution that occurs gives rise to entropy production, even when $D\to\infty$. The mechanism that drives it is not viscous dissipation, but Joule heating through charge diffusion.

This gives us the opportunity to explore a different mechanism for entropy production, and also provides a simpler set up where we can confirm the general picture that we have developed in the previous sections. It will be easy, and interesting, to consider asymmetric collisions, where the two initial black holes have different parameters, in particular different charge-to-mass ratios.

\subsection{Entropy generation in charged fusion and fission}

Our discussion will be succint, and for more details we refer to \cite{Emparan:2016sjk} and \cite{Andrade:2018rcx}. The effective theory for a charged black brane has as its variables, besides the mass density and the velocity, the charge density $q(t,\mathbf{x})$. In terms of these, the entropy density is
\beq
s=2\pi \lp m+\sqrt{m^2-2q^2}\rp \,.
\eeq
The chemical potential, conjugate to the charge, and the temperature are
\beq
\mu=\frac{2q}{m+\sqrt{m^2-2q^2}}\,,\qquad T=\frac1{2\pi}\frac{\sqrt{m^2-2q^2}}{ m+\sqrt{m^2-2q^2}}\,.
\eeq
The effective equations then imply that 
\beq
\partial_t s +\partial_i j_s^i=\kappa_q \partial_i \lp\frac{\mu}{T}\rp \partial^i \lp\frac{\mu}{T}\rp \,,
\eeq
where the expressions for the entropy current $j_s^i$ and the charge diffusion coefficient $\kappa_q$ can be found in  \cite{Emparan:2016sjk}.
The term on the right generates entropy when there is a gradient of 
\beq
\frac{\mu}{T}=4\pi\frac{q/m}{\sqrt{1-2(q/m)^2}}\,,
\eeq
that is, when $q/m$ is not homogeneous so there can be charge diffusion. Observe that, unlike in the neutral case, the temperature need not be uniform: it is smaller where $|q/m|$ is larger.

The effective equations admit exact solutions for charged blobs that are easy extensions of the neutral ones, in particular charged rotating black holes and black bars \cite{Andrade:2018rcx}. The former are the large $D$ limit of the Kerr-Newman (KN) black hole.

The entropy of a KN black hole or black bar at large $D$ is (cf.~\eqref{SMQ})
\beq
S=2\pi M \lp 1+\sqrt{1-2\fq^2}\rp\,,
\eeq
where we introduce the charge-to-mass ratio of the black hole,
\beq\label{fracq}
\fq=\frac{Q}{M}\,.
\eeq
Observe that, in this limit of $D\to\infty$, the entropy is independent of the spin. The KN black hole and the charged black bar differ in how the spin is related to the mass and charge, but are entropically equivalent.

Consider now a configuration of two KN black holes, labelled $1$ and $2$, with masses $M_{1,2}$ and charges $Q_{1,2}$. We want to study the total entropy of the system for fixed total mass $M=M_1+M_2=1$ and fixed total charge $Q=Q_1 + Q_2$. For the two remaining free parameters in the system, we use $\Delta M$ and $\Delta \fq$, such that
\beq
M_{1,2}=\frac12\pm \Delta M\,,
\eeq
\beq
\fq_{1,2} =Q-\lp\Delta M \mp \frac12\rp\Delta\fq\,,
\eeq
\ie
\beq
\fq_1-\fq_2=\Delta\fq
\eeq
where
\beq
\fq_{1,2} =\frac{Q_{1,2}}{M_{1,2}}\,.
\eeq

The entropy of the two-black hole system is
\beq
\frac{S_{(2)}}{2\pi}=\lp \frac12+\Delta M\rp\lp 1+\sqrt{1-2\fq_1^2}\rp+\lp \frac12-\Delta M\rp\lp 1+\sqrt{1-2\fq_2^2}\rp\,.
\eeq
We now ask what values of $\Delta M$ and $\Delta \fq$ maximize this entropy for fixed total $Q$ and total $M=1$. The answer is that the maximum is reached for
\beq
\Delta\fq =0
\eeq
for any value of $\Delta M$, and this maximum is equal to
\beq
\frac{S_{(0)}}{2\pi}=1+\sqrt{1-2Q^2}\,,
\eeq
which is the entropy of a single black hole or black bar with mass $M=1$ and charge $Q$ (so $\fq=Q$). That is, a system of two black holes, with possibly different masses and charges but both of them having the same charge-to-mass ratio $\fq$, has the same entropy as a single black hole or black bar with that same value of $\fq$; and a system of two black holes with different charge-to-mass ratios has lower entropy than a single black hole or black bar of the same total mass and charge.

The consequences of this for processes of black hole fusion and fission are then clear:
\begin{itemize}
\item Fission processes where an unstable black hole or black bar decays into two black holes are isentropic (adiabatic), and the final black holes will have the same charge-to-mass ratios $\fq$ as the initial one.

\item When two black holes with the same values of $\fq$ (but possibly different masses and charges) collide and merge, the subsequent process will necessarily be isentropic, regardless of whether a long-lived intermediate state forms or not, and (if there is fission) regardless of what the final outgoing black holes are. 

\item When two black holes with different values of $\fq$ collide and merge, entropy is produced. If a stable black hole forms, then it will definitely have more entropy than the initial states. The entropy production will be given by $S_{(0)}-S_{(2)}$ above.

\item If the final state consists of two outgoing black holes, then more entropy will be produced the closer the intermediate state is to a single stationary black hole or black bar (\ie the longer-lived the intermediate state is, so there is time to diffuse charge uniformly to maximize the entropy). In that case, entropy will reach a value close to saturation during the intermediate phase, with little entropy production in the subsequent fission stage.

\item If there is no long-lived, almost stationary intermediate state, then entropy production will be less than maximal, but we still expect that it will happen mostly during the fusion process (where the charge-to-mass-ratios are more different) and less so in the fission.
\end{itemize}

\begin{figure}[t]
\centering
\includegraphics[width=1\linewidth]{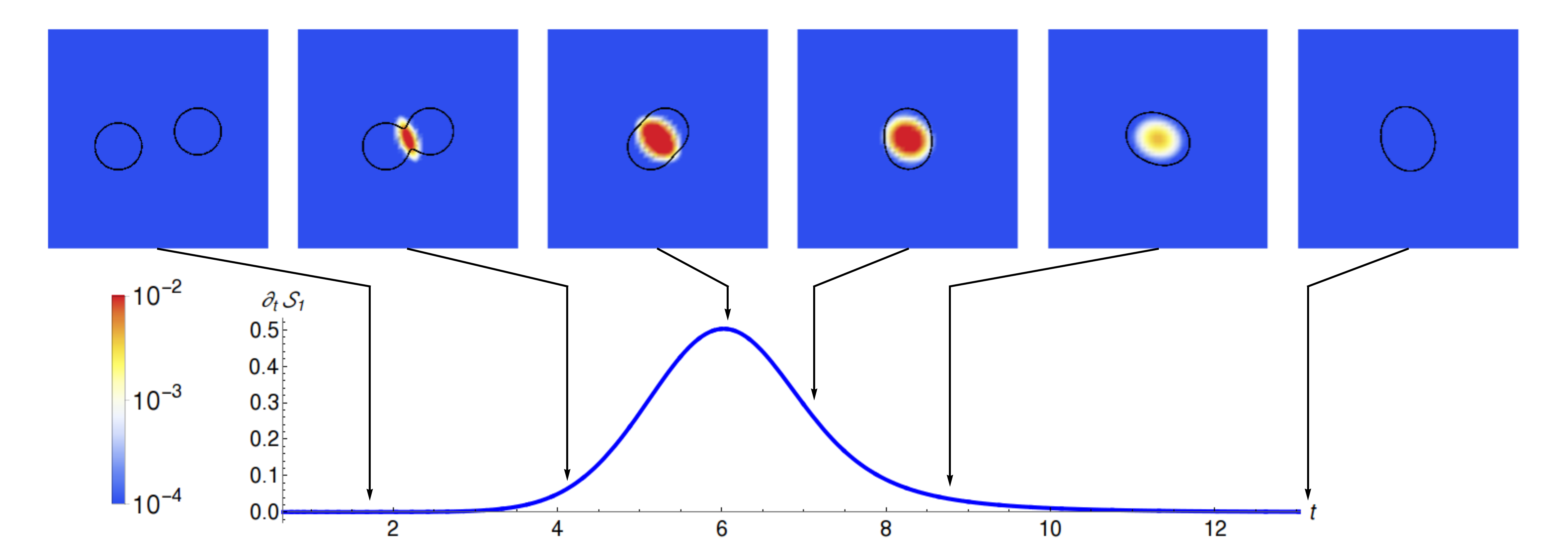}
\caption{\small Entropy production during a charged collision with fusion into a stable black bar. At LO in the effective theory, entropy is generated through charge diffusion only. The initial parameters are $q_1 = - q_2 = -0.6$, $a_1 = a_2 = 0.4$, $u_1 = - u_2 = 1$, $b = 2$, with total mass $M=1$.
\label{fig:charged}}
\end{figure}

These features (to LO at large $D$) are similar to what we have found in the neutral case (at NLO), but with stronger suppression of entropy generation in fission compared to fusion. It is easy to run numerical simulations of collisions that confirm this picture, but since conservation laws constrain much more the phenomena, they are less illustrative than in the neutral case. Therefore we only show one example of the fusion of two oppositely charged black holes, fig.~\ref{fig:charged}. It is qualitatively similar to fig.~\ref{fig:EntropyFrames}, even in its duration.

\section{Final comments and outlook}\label{sec:concl}

Our study has produced a consistent picture of the phenomena of fusion and fission of black holes, including the entropy generation mechanisms at different stages. One of the main results has been to highlight the attractor role that the intermediate, long-lived, quasi-thermalized black bar phase plays in a $2\to 1\to 2$ collision with fission, and how it is connected to a principle of total entropy maximization. 

It is surprising that entropy maximization is somehow driving the dynamics. The system obeys time-irreversible equations that imply that, if at some moment in the evolution entropy is generated, then there is no coming back. But, in principle, the dynamics does not force the system, at least not in any manifest way, to evolve in a direction where entropy grows---it only forbids it to decrease---and even less so that entropy should grow as much as possible compatibly with conservation laws. The surprise that we find is that the evolution does lead to an end state very close to maximum entropy among a continuum of kinematically and geometrically allowed states. In statistical and quantum mechanics systems evolve stochastically sampling nearby configurations---and then, final thermodynamic equilibrium is achieved when entropy reaches a maximum. Here, however, we have a completely deterministic classical system whose equations seem to drive it in a direction that almost maximizes the rather non-obvious quantity $S_1$. The time scale involved is much shorter than, say, the scrambling time for a black hole (which, in the strict classical limit, is an infinite time). And moreover, for all we can see, entropy is in general almost, but not quite, maximized among possible final states. Indeed, by adjusting the initial conditions (e.g., to make a black string break up into multiple static black holes, or colliding black holes with very large $J/M$) the difference to the maximum can be made larger. Maximal entropy provides only an approximate criterion, but a remarkably accurate and powerful one.

In addition, we have also produced a detailed temporal and spatial tomography of entropy generation.  Fusion, as might be expected, is highly irreversible.  The production of entropy during fusion is relatively featureless, characterized by an initial peak in the production rate, where both shear and bulk viscosity contribute. If the system then settles down to a (long-lived) bar, a second stage with smaller entropy production can appear. Fission of unstable configurations follows the pattern of the decay of unstable black strings, with a duration of the order of \eqref{tfission}, and approximately equal amounts of dissipation of shear and expansion.

Our results are expected to be most applicable for black hole evolution in $D\geq 6$, but we would like to elaborate more on their possible qualitative relevance in four dimensions. The most important differences between $D=4$ and $D\geq 6$\footnote{The case $D=5$ is in some respects closer to $D=4$ and in others to $D\geq 6$. We will not refer to its peculiarities here.} refer to the dynamics of rotation: (i) quasi-stable Keplerian orbits (in General Relativity they are not fully stable due to gravitational wave emission), which are important in the evolution towards a merger, are possible in $D=4$ but not in $D\geq 6$; (ii) the properties of rotating black holes differ markedly in the two cases: in $D=4$ their spin is bounded above, while in $D\geq 6$ it is unbounded. Moreover, in $D=4$ the Kerr black hole (and not, \eg a black bar) will always be the endpoint of fusion, without any instability that would lead to its fission. If the total angular momentum in the system is above the Kerr bound for the final black hole, then the excess will be shed off into radiation, possibly involving an `orbital hang up' stage that delays the merger \cite{Campanelli:2006uy,Baumgarte:2010ndz}. In $D\geq 6$ instead, the upper bound on the angular momentum is not absolute but dynamical and set by an instability, so the excess angular momentum does not result in hang up but triggers fission.

We do not expect our studies of fission, nor of the relaxation to a stable black bar, to have application to collisions between Kerr black holes. But  when studying fusion into a stable rotating MP black hole, the differences we have mentioned are less important than they may appear. The reason is that in $D=4$ the final plunge before two black holes merge occurs when their orbit becomes unstable. And when the spin of MP black holes is below the ultraspinning bound, their properties are similar to the Kerr black hole.  It is interesting that in our simulations of collisions with relatively large impact parameters, we have observed that, prior to coalescence, the black hole trajectories are deflected into what looks like an approximately circular orbit (unstable dumbbells appear to describe such configurations), until the two black holes finally plunge towards each other. Qualitatively at least, this resembles the four-dimensional evolution.

So, as long as the angular momenta involved are moderate, the dynamics of black hole collisions and mergers in $D\geq 6$ are qualitatively similar to $D=4$, and our study of how entropy is produced (with the caveats that concern gravitational wave emission) should provide at least a guide to what to expect in that case.

\section*{Acknowledgments}

Work supported by ERC Advanced Grant GravBHs-692951, MEC grant FPA2016-76005-C2-2-P, and AGAUR grant 2017-SGR 754. 
RL is supported by the Spanish Ministerio de Ciencia, Innovaci\'on y Universidades Grant FPU15/01414.
RS is supported by JSPS KAKENHI Grant Number JP18K13541 and partly by Osaka City University Advanced Mathematical Institute (MEXT Joint Usage/Research Center on Mathematics and Theoretical Physics).

\newpage

\appendix

\section{Effective theory entropy of AdS black branes}\label{app:AdS}

The extension of the study in sec.~\ref{sec:efftheory} to black branes in AdS only needs a few sign changes. We mark them with the parameter
\beq\label{AFAdS}
\epsilon=\begin{cases}+1
&\mathrm{for~AF}\,,\\ -1 &\mathrm{for~AdS}\,.
\end{cases}
\eeq
This sign difference is responsible for the different stability properties of AF and AdS black branes. 

The effective equations are the same as \eqref{dtm} and \eqref{dtmv} but now
with
\beq\label{stressneuteps}
\tau_{ij}= - \epsilon\,m\,\delta_{ij}  -2m \partial_{(i}v_{j)}- m\,\partial_j\partial_i \ln m\,.
\eeq
The NLO entropy density is
\beq\label{s1eps}
s_1=4\pi\lp -\frac12 m v_i v^i -\frac1{2m}\partial_i m\,\partial^i m +\epsilon\, m \ln m\rp\,,
\eeq
and the form of the entropy current \eqref{j1} remains the same, with the appropriate $s_1$ and $\tau_{ij}$. 
Then, the non-negative divergence of the entropy current is satisfied in the same form as in \eqref{dts1}, and \eqref{nloent2} extends to
\beq\label{nloent2eps}
s(t,x)=4\pi \lp \bar{m}-\frac1{D} \lp \frac12 m v^2 +\frac1{2m}(\partial m)^2-c_s\rp\rp^{1+\epsilon/D}\,.
\eeq

Notice that for the AdS black brane (whose worldvolume is infinite dimensional, even if only a finite number of directions are active),
\beq
\zeta=0
\eeq
due to conformal invariance, which implies that viscosity does not dissipate expansion.

\section{Stress-energy and entropy in the effective theory to $1/D$}
\label{app:nlo}

Let us assume the $D=n+p+3$ spacetime is written in Eddington-Finkelstein coordinates
\begin{eqnarray}
 ds^2 = - A dt^2 + 2 u_t dt dr -2C_I dt dZ^I+ H_{IJ}dZ^I dZ^J.
\end{eqnarray}
For the spatial metric, we assume a rescaling in the `active' dimensions and spherical symmetry in the `passive' ones,
\begin{eqnarray}
H_{IJ}dZ^I dZ^J =  \fr{n}G_{ij}dz^i dz^j + r^2 d\Omega_{n+1}^2,\label{eq:phys-resc-sph}
\end{eqnarray}
where $G_{ij}$ is the $p$-dimensional metric.

The  metric is expanded in $1/n$,
\begin{equation}
 A= \sum_{k=0} \frac{ A_{[k]}}{n^k},\quad u_t = 1 + \sum_{k=0} \frac{u_{t,[k]}}{n^{k+1}},\quad C_i =  \sum_{k=0} \frac{C_{i,[k]}}{n^{k+1/2}},\quad G_{ij} = \delta_{ij} + \sum_{k=0} \frac{G_{ij,[k]}}{n^{k+1}},
\end{equation}
where we introduce a radial coordinate ${\sf R} = r^n$.
At leading order, we obtain
\begin{equation}
 A_{[0]} = 1- \frac{m}{{\sf R}},\quad C_{i,[0]} = \frac{p_i}{\sf R},
 \quad G_{ij,[0]} = \frac{p_i p_j}{m \sf R}, \quad
 u_{t,[0]} = -\frac{p_i p^i}{2 m {\sf R}^2},
\end{equation}
where $m=m(t,z)$ and $p_i=p_i(t,z)$ are integration functions. To avoid ambiguity in the definition of the integration functions at higher order, we fix $m$ and $p_i$ by
\begin{equation}
 A({\sf R}=m) = 0 ,\quad C_i ({\sf R}=m) = \frac{p_i}{m}.
\end{equation}
Note that in general ${\sf R}=m$ differs from the horizon position, which will be relevant below.

\subsection{Quasi-local stress tensor}
The quasi-local stress tensor is defined at the asymptotic boundary of the near-horizon region,
\begin{equation}
{\bf T}_{\mu\nu} = \lim_{r\to\infty}\frac{\Omega_{n+1} r^{n+1}}{8\pi G}(K \gamma_{\mu\nu} - K_{\mu\nu})+({\rm regulator})\,,
\end{equation}
where $(\gamma_{\mu\nu},K_{\mu\nu})$ are the metric and extrinsic curvature on a  surface at constant $r$. The regulator terms are chosen to eliminate the divergent terms at $r\to\infty$.
The boundary metric is given by
\begin{equation}
 ds^2 = - dt^2 + \fr{n} dz^i dz_i\,. \label{eq:boudary-metric}
\end{equation}
For convenience, we  use the dimensionless tensor,
\begin{equation}
 {\bf T}_{\mu\nu} = \frac{(n+1)\Omega_{n+1}}{16\pi G}T_{\mu\nu}.
\end{equation}
The result up to NLO in the $1/n$ expansion is given by\footnote{Here the indices of $T_{\mu\nu}$ are raised with the boundary metric~(\ref{eq:boudary-metric}), while $p_i$ and $\partial_i$ with $\delta^{ij}$.}
\begin{align}
&T^{tt} = m - \fr{n}(2+\ln m)\partial_i p^i,\label{Ttt}\\
&T^{ti}= p_i - \partial_i m - \fr{2nm^2} \bigl[
2m(m+\partial_i p^i)(p^i-\partial^i m)+2p^i p^j \partial_j m- p_j p^j\partial^i m
+4 m p_j \partial^{[j} p^{i]}\nonum
&\hspace{4cm} +\left(2m\partial_j (p^ip^j)-2p^i p^j \partial_j m
-2m^2 \partial^i \partial_j p^j\right) \ln m \bigr]\,,\\
& T^{ij} = \alpha \delta^{ij} + \beta_{ij}\,,
\end{align}
where
\begin{align}
&\alpha = \left(1-\fr{n}\right)(-m+\partial_t m + \partial_k p^k)\nonum
&\quad + \fr{n}\left[ -\fr{2m}\partial^2 (p^2) - \partial_k p^k \left(1-\frac{2p^2}{m^2} + \frac{3p^k \partial_k m}{m^2}\right)
 - \frac{p^k-\partial^k m}{m}\partial_k (p^2) + \frac{p^k}{m^2}\partial^\ell m \partial_k p_\ell  
 \right. \nonum
 &\quad \left.+ \frac{p^2}{m} \left(1+\frac{\partial^2 m}{2m}\right)  - \partial^2 \left(\frac{p^2}{m}\right) \ln m+\partial_k p^k \ln m -\partial_t ( \partial_k p^k \ln m)\right],
\end{align}
and
\begin{align}
&\beta_{ij} =\left(1-\fr{n}\right) \left( - 2 \partial_{(i} p_{j)} + \frac{p_i p_j}{m}\right)\nonum
&\quad + \fr{n}\left[- \frac{p^2}{m^3} \partial_i m\partial_j m + \left(\frac{2p^k \partial_k m}{m^2}-\frac{(\partial m)^2}{m^2}-\frac{3\partial_k p^k}{m}\right)\frac{p_i p_j}{m}\right.
 \nonum
 &\quad + \frac{2}{m^3}(-p^2+2m\partial_k p^k) p_{(i}\partial_{j)} m
  + \left(\frac{2p^k\partial_k m}{m^2}-\frac{p^2}{m^2}-2\ln m\right)\partial_{(i} p_{j)}\nonum
  &\quad \left. +\frac{2}{m^2} \left(p_{(i}\partial_{j)}(p^2) - \partial_k m p_{(i} \partial_{j)} p^k\right) + \partial_i \partial_j \left(\frac{p^2}{m}\right)\ln m+(-1+\ln m) \partial_t \left(\frac{p_i p_j}{m}\right) \right].
\end{align}
Here we have written $p^2 = p_i p^i$ and $\partial^2 = \partial_k \partial^k$.

\subsection{Entropy density and entropy current}\label{app:nloent}

The position of the event horizon $\Phi = r-r_h(t,Z^I)=0$ is given by the null condition for the normal vector $d\Phi = dr-\partial_tr_h dt -\partial_I r_h dZ^J$,
\begin{eqnarray}
d\Phi^2= \fr{u_t^2}\left[A-2u_t\partial_t r_h+H^{IJ}(C_I-u_t \partial_J r_h)(C_J-u_t \partial_J r_h) \right]=0.\label{eq:null-hs-eh}
\end{eqnarray}
In the dynamical case, this condition does not give the actual event horizon but rather the local one. However, as in \cite{Bhattacharyya:2008xc}, this is useful to define the entropy current on the black brane.
Expanding up to NLO in $1/n$  we obtain
\begin{equation}
 {\sf R}_h=r_h^n = m - \fr{n}\left(\frac{p^ip_i}{m} -2 p^i\partial_i \ln m + \fr{m}\partial_i m \partial^i m-2\partial_t m\right).
\end{equation}
One can see that $\cR = m$ also gives the event horizon for the static solution. With a rigid rotation, however, we have ${\sf R}_h \neq m$ beyond LO.
Using eq.~(\ref{eq:null-hs-eh}), the geometry on the horizon becomes
\begin{eqnarray}
\left. ds^2\right|_{\cal H} =
 h_{IJ} (dZ^I-V^I dt)( dZ^J-V^Jdt),
\end{eqnarray}
where $h_{IJ}=\left. H_{IJ} \right|_{\cal H}$ and $V^I =h^{IJ}( \left. C_J-u_t \partial_J r_h\right|_{\cal H})$.
Following~\cite{Bhattacharyya:2008xc}, the entropy $(D-2)$-form is defined from the area-form of this surface,
\begin{equation}
 {\cal A} = \fr{4G} \sqrt{h} (dZ^1-V^1 dt) \wedge \dots \wedge(dZ^{D-2}-V^{D-2} dt),
\end{equation}
which determines the entropy current to be
\begin{equation}
 {\cal A} = \frac{\epsilon_{{\bar\mu} {\bar\mu}_1\dots {\bar\mu}_{D-2}}}{(D-2)!}\bar{\cal J}_s^{\bar\mu}   dX^{{\bar\mu}_1}
\wedge \dots \wedge dX^{{\bar\mu}_{D-2}},
\end{equation}
where $X^{\bar\mu} = (t,Z^I)$. By comparison, one obtains
\begin{eqnarray}
 \bar{\cal J}_s^{\bar\mu} \partial_{\bar\mu} = \fr{4G}\left(\sqrt{h}\partial_t+ \sqrt{h}V^I\partial_I\right).
\end{eqnarray}
Recalling the spatial setup~(\ref{eq:phys-resc-sph}), the entropy current reduces to
\begin{eqnarray}
 {\cal J}_s^\mu \partial_\mu = \frac{\Omega_{n+1}r_h^{n+1}}{4G n^{p/2}}\left(\sqrt{\cal G}\partial_t+ \sqrt{\cal G}V^i\partial_i\right),
\end{eqnarray}
where $x^\mu = (t,z^i)$ and ${\cal G}_{ij}=\left. G_{ij}\right|_{\cal H}$.
The dimensionless version is
\begin{equation}
{\cal J}_s^\mu = \frac{\Omega_{n+1}}{4G n^{p/2}}J_s^\mu.
\end{equation}
For the black brane, the result up to NLO in $1/n$ expansion is
\begin{align}
& J^t = m + \fr{2nm}(2 m^2\ln m - p^2 + 4 p^i \partial_i m-2 (\partial m )^2+4 m \partial_t m)\,,\\
& J^i = p^i - \partial^im
 + \fr{2nm^2}\bigr(-p^2 p^i 
+4m (p_j-\partial_j m) (\partial^i p^j-\partial^i \partial^j m)
- 4 m^2 \partial_t \partial^i m\nonum
& \hspace{5cm} +2m^2\ln m(p^i-2 \partial^i m) 
\bigl).
\end{align}

We can use the LO equations \eqref{dtm} and \eqref{dtmv}
to find that in the effective theory the entropy density is
\beqa
s&=&4\pi J^t\nn\\
&=&4\pi \bar{m}+\frac{4\pi}{n}\lp (m+\partial_i p^i)\ln m+2\partial^2 m -\frac1{2m}(p^2-4p^i\partial_i m+2\partial_i m\partial^i m)\rp\,,\label{seff1}
\eeqa
where we have defined the mass density to NLO from \eqref{Ttt}
\beq
\bar{m}=m-\frac1{n}\lp 2+\ln m\rp \partial_i p^i\,.
\eeq
Using again the LO equations with
\beq
p_i=m v_i +\partial_i m\,,
\eeq
we can rewrite \eqref{seff1} as 
\beq
s=4\pi \bar{m}+\frac{4\pi}{n}\lp -\frac12 m v_i v^i-\frac1{2m}(\partial m)^2+m\ln m\rp+\text{total divergence}\,.
\eeq
Dropping the total divergence term, which does not contribute to the integrated entropy, we obtain  \eqref{s1} and \eqref{nloent}.

\section{Boost invariance of the entropy}\label{app:boost}

The effective theory is invariant under Galilean symmetry. Therefore a Galilean transformation acting on a blob solution $m(t,x),\, v_i(t,x)$ yields another solution
\begin{equation}
m'(t,x) = m(t,x-X(t)),\qquad  v'_i(t,x) = v_i(t,x-X(t))+u_i
\end{equation}
where $X_{i}(t)=u_i t + b_i$.
The mass and the linear and angular momenta transform as
\begin{equation}
 M' = M,\qquad P'_i = P_i+M u_i \qquad J'_{ij} =J_{ij} + (b_i u_j-b_j u_i)M\,.
\end{equation}
The first two  are actually the Lorentz transformation (setting for simplicity $u_i = (u,0,\dots,0)$)
\begin{equation}
 \left(\begin{array}{c}{M'}\\ \fr{\sqrt{D}}{P'}_x\\ \fr{\sqrt{D}}{P'}_{i\neq x} \end{array}\right) = 
 \left(\begin{array}{ccc}\cosh\alpha & \sinh\alpha &0\\ \sinh\alpha & \cosh\alpha &0\\ 0&0&\delta_{ij}\end{array}\right) \left(\begin{array}{c}M\\ \fr{\sqrt{D}}P_x\\ \fr{\sqrt{D}}P_{j\neq x}\end{array}\right),\label{eq:traveling-blob-trans}
\end{equation}
up to leading order in the large-$D$ limit of non-relativistic velocities,
\begin{equation}
 \alpha = \textrm{arctanh} \frac{u}{\sqrt{D}}=\frac{u}{\sqrt{D}}+\ord{\frac1{D^{3/2}}}.
\end{equation}

Although the masses remain invariant in the LO effective theory, the Lorentz transformation generates terms at NLO
\begin{equation}
 \left(\begin{array}{c}{M'}\\ \fr{\sqrt{D}}{P'}_x\\ \fr{\sqrt{D}}{P'}_{i\neq x} \end{array}\right) = 
 \left(\begin{array}{ccc}1+\frac{u^2}{2D} & \frac{u}{\sqrt{D}} &0\\ \frac{u}{\sqrt{D}} & 1+\frac{u^2}{2D} &0\\ 0&0&\delta_{ij}\end{array}\right) \left(\begin{array}{c}M\\ \fr{\sqrt{D}}P_x\\ \fr{\sqrt{D}}P_{j\neq x}\end{array}\right),\label{eq:traveling-blob-trans2}
\end{equation}
whose effect we must take into account when computing the Lorentz transformation of the entropy.

The NLO entropy \eqref{S1} transforms as
\begin{align}
 &S_1'= 4\pi \int_{{\mathbb R}^p} d^px \left(-\fr{2}m'{v'}^2 - \fr{2m'} (\partial m')^2+ m' \log m'\right)\nonum
&=4\pi  \int_{{\mathbb R}^p} d^px \left(-\fr{2}mv^2 - \fr{2m} (\partial m)^2 + m \log m-mv_i u^i - \fr{2} m u^2\right)\nonum
 &  = S_1 - 4\pi\lp u^i P_i + \frac{1}{2}u^2 M\rp\,,
 \end{align}
which implies that the mass-normalized entropy \eqref{calS1} transforms as
\begin{equation}
\mc{S}_1' =  \mc{S}_1- \fr{2} u^2-\frac{u^i P_i}{M}\,.
\end{equation}
We see that the NLO entropies are not boost invariant. However, it is straightforward to verify that under \eqref{eq:traveling-blob-trans2} the total entropy
\begin{equation}
 S= 4\pi M + \fr{D}S_1=  4\pi M' + \fr{D}S_1'=S'
\end{equation}
is invariant. The mass-normalized total entropy \eqref{physSeftS1} is not, since the mass (energy) is not boost invariant.

If the `unprimed' frame is at rest, so that $P_i =0$ and $M$ is the rest mass, then
\begin{equation}
 M' = \frac{M}{\sqrt{1-\frac{u^2}{D}}}= M+\frac1D\frac{M u^2}{2}+\ord{\frac1{D^2}}\,
\end{equation}
and 
\beq\label{S1boost}
\mc{S}_1'=\mc{S}_1- \fr{2} u^2\,.
\eeq
This directly yields \eqref{S1MP} from \eqref{calS1MP}.

\section{Physical magnitudes}\label{app:phys}

Following \cite{Andrade:2019edf}, the physical mass, entropy, and spin (boldfaced) are given in terms of the effective theory magnitudes as\footnote{We correct a typo in \cite{Andrade:2019edf}, where $\Omega_{D-4}\to \Omega_{D-2}$.}
\beqa
\mathbf{M}&=&\frac{\Omega_{D-2}}{16\pi G} r_+^{D-3}\,\frac{D-2}{2\pi m_0} M\,,\\
\mathbf{S}&=&\frac{\Omega_{D-2}}{16\pi G} r_+^{D-2}\,\frac{1}{2\pi m_0} S\,,\\
\mathbf{J}&=&\frac{2 r_+}{D-2}\frac{J}{M}\mathbf{M}\,.
\eeqa
Here $r_+$ is a length scale that is invariantly defined as the horizon radius of the transverse sphere $S^{D-4}$ at the rotation axis, and $m_0$ is a dimensionless parameter that relates it to the physical temperature $\mathbf{T}$, 
\beq
m_0=\lp\frac{4\pi}{D-5}\mathbf{T}r_+\rp^{D-5}\,.
\eeq
From these expressions, we obtain the mass-normalized, dimensionless entropy \eqref{mcS} as
\beqa
\mc{S}&=&\frac{S}{4\pi M}\lp 1-\frac1{D}\ln\frac{M}{2\pi m_0}\rp\nn\\
&=& 1+\frac1{D}\lp c_s+\ln 2\pi m_0+\mc{S}_1\rp\,,\label{calSS1}
\eeqa
where in the last expression we have used \eqref{SS1} and \eqref{calS1}.

In order to determine the constant $c_s$ we apply these formulas to the MP blob solution \eqref{MPblob} so that we recover the known results for the exact MP black hole solution. These are
\beqa
\mathbf{M}&=&\frac{\Omega_{D-2}}{16\pi G}(D-2)r_m^{D-3}\,,\\
\mathbf{S}&=&\frac{\Omega_{D-2}}{4 G}r_m^{D-3} r_+\,,
\eeqa
where the mass-radius $r_m$ and the horizon radius $r_+$ are related by
\beq
\frac{r_+}{r_m}=\lp 1+\frac{a^2}{r_+^2}\rp^{-1/(D-3)}\simeq
1-\frac1{D}\ln\lp 1+\frac{a^2}{r_+^2}\rp\,.
\eeq
Then, for this black hole, the mass-normalized, dimensionless entropy defined in \eqref{mcS} is
\beq\label{mcSMP}
\mc{S}= 1-\frac1{D}\ln(1+a^2)\,,
\eeq
where, henceforth, since we work with scale-invariant quantities, we can set $r_+=1$.

For the MP blob solution \eqref{MPblob}, the effective theory NLO entropy \eqref{S1} is
\beq
\frac{S_1}{4\pi M}=-\frac{2a^2}{1+a^2}\,,
\eeq
and then the mass-normalized one, $\mc{S}_1$ in \eqref{calS1},  is given in \eqref{calS1MP}.
Plugging the latter into \eqref{calSS1} and setting
\beq
c_s=2-\ln m_0\,,
\eeq
we recover correctly the physical value \eqref{mcSMP}. Using this now in the general formula \eqref{calSS1}, we obtain \eqref{physSeftS1}.

\section{Geometric constraint on $b_\mathrm{out}$}\label{app:bout}

The most sensible and robust way of relating $b_\mathrm{out}$ to other parameters in the collision is to demand that it be not larger than twice the radius of the two outgoing black holes
\beq\label{boutR}
b_\mathrm{out} \lesssim 2 R_\mathrm{out}\,.
\eeq
This is naturally expected from the fission of the intermediate blob; any residual late interaction between the outgoing blobs after fission (\eg a thin tube stretching between them) will be attractive and tend reduce the impact parameter. Furthermore, since entropy maximization drives $b_\mathrm{out}$ to the largest possible values, we expect that the actual value of $b_\mathrm{out}$ is close to saturating \eqref{boutR}.

While this criterion is indeed reasonable, its implementation in the effective theory faces the problem that the radius of the black hole as a gaussian blob such as \eqref{MPblob} is not clearly defined; it requires choosing a specific boundary radius at which the blob density $m(r)$ is only a certain fraction of what it is at its center. That is, we define the boundary radius of the blob, $R$, in terms of a small number $\epsilon_b$ as
\beq\label{meps}
m(R)=\epsilon_b\, m_0\,.
\eeq
The choice of $\epsilon_b$ is somewhat arbitrary but since
\beq
R=\sqrt{2(1+a^2)\ln\epsilon_b^{-1}}\,,
\eeq
the dependence on $\epsilon_b$ is doubly suppressed by the logarithm and the square root. 
\begin{figure}[t]
\centering
\includegraphics[width= 0.6\linewidth]{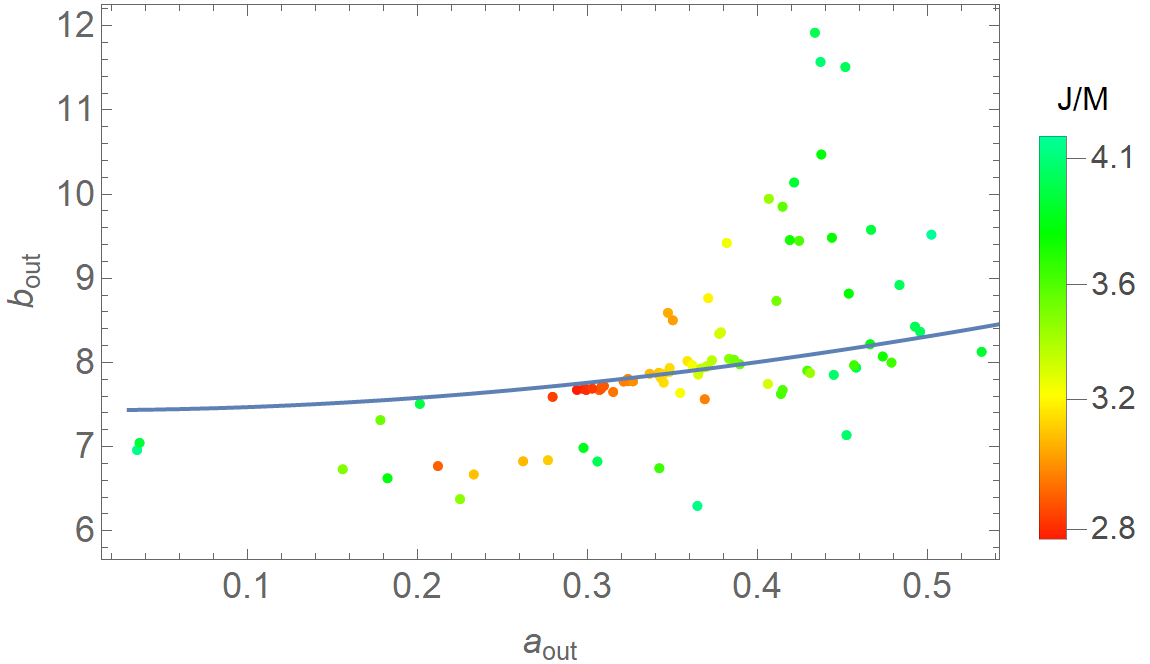}
\caption{\small Outgoing parameters $(a_{\text{out}}, b_{\text{out}})$ for the collisions in fig.~\ref{fig:innout} (dots), and for the geometric estimate \eqref{boutguess} (blue curve).
\label{fig:baout}}
\end{figure}
These arguments give \eqref{boutguess}.
For collisions near the critical attractor, our numerical simulations give $b_\mathrm{out}\approx 7.8$, $a_\mathrm{out}\approx 0.3$, which are well reproduced for
\beq
\epsilon_b\approx 10^{-3}\,.
\eeq
Of course this is a fit to a point, but \eqref{boutguess} also predicts that $b_\mathrm{out}$ grows with $a_\mathrm{out}$. As shown in fig.~\ref{fig:baout}, the dependence is in good agreement with the numerical data when the value of $J/M$ is not very high. At large $J/M$, where the dispersion in the data is higher, the intermediate blob is very elongated, which can lead to \eqref{boutguess} either underestimating $b_\mathrm{out}$ (the blobs are far apart when they split) or overestimating it (due to late-time attraction caused by a thin tube between the blobs). It would be interesting to better understand these effects.

\end{document}